\documentclass[aps,pra,reprint,superscriptaddress,showpacs]{revtex4-1}
\usepackage{amsmath}
\usepackage{amsfonts}
\usepackage{amsthm}
\usepackage{amssymb}
\usepackage{graphicx}
\usepackage{enumerate}
\usepackage{color}
\usepackage[T1]{fontenc}
\usepackage{mathptmx}
\usepackage{braket}
\usepackage{todonotes}
\usepackage{soul}
\usepackage{subfigure}
\graphicspath{{figures/}}

\usepackage[bookmarks=false,colorlinks,citecolor=blue,
linkcolor=blue,anchorcolor=blue,urlcolor=blue
]{hyperref}

\renewcommand{\url}[1]{}
\newcommand{\urlprefix}{}
\renewcommand{\href}[1]{}

\begin{document}
\title{Phase-Matching Quantum Cryptographic Conferencing }
\author{Shuai Zhao}
\affiliation{Hefei National Laboratory for Physical Sciences at Microscale and Department of Modern Physics, University of Science and Technology of China, Hefei, Anhui 230026, People's Republic of China\\}
\affiliation{CAS Center for Excellence and Synergetic Innovation Center of Quantum Information and Quantum Physics, University of Science and Technology of China, Hefei, Anhui 230026, People's Republic of China\\}
\author{Pei Zeng}
\affiliation{Center for Quantum Information, Institute for Interdisciplinary Information Sciences, Tsinghua University, Beijing 100084, People's Republic of China\\}
\author{Wen-Fei Cao}
\affiliation{Hefei National Laboratory for Physical Sciences at Microscale and Department of Modern Physics, University of Science and Technology of China, Hefei, Anhui 230026, People's Republic of China\\}
\affiliation{CAS Center for Excellence and Synergetic Innovation Center of Quantum Information and Quantum Physics, University of Science and Technology of China, Hefei, Anhui 230026, People's Republic of China\\}
\author{Xin-Yu Xu}
\affiliation{Hefei National Laboratory for Physical Sciences at Microscale and Department of Modern Physics, University of Science and Technology of China, Hefei, Anhui 230026, People's Republic of China\\}
\affiliation{CAS Center for Excellence and Synergetic Innovation Center of Quantum Information and Quantum Physics, University of Science and Technology of China, Hefei, Anhui 230026, People's Republic of China\\}
\author{Yi-Zheng Zhen}
\affiliation{Institute for Quantum Science and Engineering, Southern University of Science and Technology, Shenzhen, Guangdong 518055, People's Republic of China\\}
\affiliation{Hefei National Laboratory for Physical Sciences at Microscale and Department of Modern Physics, University of Science and Technology of China, Hefei, Anhui 230026, People's Republic of China\\}
\affiliation{CAS Center for Excellence and Synergetic Innovation Center of Quantum Information and Quantum Physics, University of Science and Technology of China, Hefei, Anhui 230026, People's Republic of China\\}

\author{Xiongfeng Ma}
\email{xma@tsinghua.edu.cn}
\affiliation{Center for Quantum Information, Institute for Interdisciplinary Information Sciences, Tsinghua University, Beijing 100084, People's Republic of China\\}
\author{Li Li}
\email{eidos@ustc.edu.cn}
\affiliation{Hefei National Laboratory for Physical Sciences at Microscale and Department of Modern Physics, University of Science and Technology of China, Hefei, Anhui 230026, People's Republic of China\\}
\affiliation{CAS Center for Excellence and Synergetic Innovation Center of Quantum Information and Quantum Physics, University of Science and Technology of China, Hefei, Anhui 230026, People's Republic of China\\}
\author{Nai-Le liu}
\email{nlliu@ustc.edu.cn}
\affiliation{Hefei National Laboratory for Physical Sciences at Microscale and Department of Modern Physics, University of Science and Technology of China, Hefei, Anhui 230026, People's Republic of China\\}
\affiliation{CAS Center for Excellence and Synergetic Innovation Center of Quantum Information and Quantum Physics, University of Science and Technology of China, Hefei, Anhui 230026, People's Republic of China\\}
\author{Kai Chen}
\email{kaichen@ustc.edu.cn}
\affiliation{Hefei National Laboratory for Physical Sciences at Microscale and Department of Modern Physics, University of Science and Technology of China, Hefei, Anhui 230026, People's Republic of China\\}
\affiliation{CAS Center for Excellence and Synergetic Innovation Center of Quantum Information and Quantum Physics, University of Science and Technology of China, Hefei, Anhui 230026, People's Republic of China\\}

\begin{abstract}
Quantum cryptographic conferencing (QCC) holds promise for distributing information-theoretic secure keys among multiple users over long distance. Limited by the fragility of Greenberger-Horne-Zeilinger (GHZ) state, QCC networks based on directly distributing GHZ states at long distance still face big challenge. Another two potential approaches are measurement device independent QCC and conference key agreement with single-photon interference, which was proposed based on the post-selection of GHZ states and the post-selection of W state, respectively. However, implementations of the former protocol are still heavily constrained by the transmission rate $\eta$ of optical channels and the complexity of the setups for post-selecting GHZ states. Meanwhile, the latter protocol cannot be cast to a measurement device independent prepare-and-measure scheme. Combining the idea of post-selecting GHZ state and recently proposed twin-field  quantum key distribution protocols, we report a QCC protocol based on weak coherent state interferences named phase-matching quantum cryptographic conferencing, which is immune to all detector side-channel attacks. The proposed protocol can improve the key generation rate from $\mathrm{O}(\eta^N)$ to $\mathrm{O}(\eta^{N-1})$ compared with the measurement device independent QCC protocols. Meanwhile, it can be easily scaled up to multiple parties due to its simple setup.
\end{abstract}

%%%%%%%%%%%%%%%%%%
%%%%%%%%%%%%%%%%%%
\maketitle

\section{Introduction}
Quantum network~\cite{Elliott_2002Building,elliott2005current,Peev_2009The,Xu2009Field,Stucki_2011Long,kimble2008quantum,Liao2018Satellite,Wehnereaam9288Quantum,caleffi2018quantum,castelvecchi2018quantum}, aimed at realizing quantum information tasks among multiple parties, is playing more and more important roles in burgeoning quantum information processing including quantum computing~\cite{Steane_1998}, quantum communication~\cite{Gisin2007} and quantum metrology~\cite{Giovannetti2006}. Quantum Cryptographic Conferencing (QCC) network~\cite{bose1998multiparticle,chen2007multi,chen2005conference,fu2015long,grasselli2019conference,murta2020quantum}, which distributes information-theoretic secure keys among multiple parties over long distance, is one of the most promising applications in quantum information science. With the rapid development of quantum information processing, QCC network is of great potential to improve the security of the communications in networks. For example, QCC network can be used to broadcast message to users securely. So far, several protocols are proposed to realize QCC networks. The first protocol is based on the predistribution of multi-party entanglement states~\cite{bose1998multiparticle,chen2007multi,chen2005conference}. These presentations require the predistribution of Greenberger-Horne-Zeilinger (GHZ) entanglement state~\cite{Greenberger1989}, which is initially introduced to verify Bell's theorem~\cite{bell1964einstein,brunner2014bell}. Though great endeavours have been made to improve preparation of multipartite GHZ states~\cite{bourennane2003multiphoton,pan2012multiphoton,monz201114,wang2016experimental,song201710,chen2017observation,wang201818}, the low intensity and fragility of the GHZ states make its applying to practical QCC network facing big challenge within current technology. The second protocol is measurement device independent QCC (MDI-QCC) which is based on the post-selection of GHZ state~\cite{fu2015long}. Once a successful detection event occurs, a GHZ state is shared among multiple parties~\cite{qian2005universal}. Thus, multiple parties can distribute secret key bits among them by the post-selected entanglement states. Further, the measurement device can be controlled by an untrusted third party, Eve. Therefore, according to the measurement device independent quantum key distribution (MDI-QKD) idea~\cite{lo2012measurement} (see also~\cite{braunstein2012side}), it is immune to all detector side-channel attacks. Combined with decoy-state method~\cite{lo2005decoy}, MDI-QCC network is promised more reasonably to be realized in experiments within current technology. The third protocol is the conference key agreement with single photon interference (single-photon CKA)~\cite{grasselli2019conference}, which is based on post-selection of the W state~\cite{dur2000three}.  However, the single-photon CKA protocol cannot be cast to a MDI prepare-and-measure scheme. Meanwhile, the signal pulses cannot be substituted by coherent states, and the local qubits have to resort to quantum memories. Thus, the feasibility of single-photon CKA requires further investigation~\cite{grasselli2019conference}.
\begin{figure}
\includegraphics[width =0.45\textwidth]{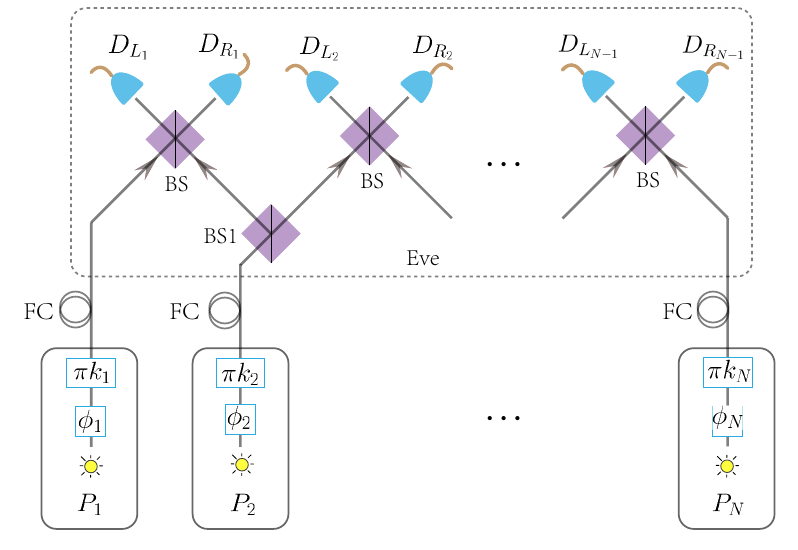}
\caption{Schematic setup for $N$-party PM-QCC network. $\phi_1$, $\phi_2$ and $\phi_N$ $\in[0,2\pi)$ label the random phases for parties $P_1$, $P_2$ and $P_N$, respectively. The $k_1$, $k_2$ and $k_N$ $\in \{0,1\}$ label the random bits for parties $P_1$, $P_2$ and $P_N$, respectively. $D_{L_1}(D_{R_1})$: the left (right) detector of the first measurement branch. $D_{L_2}(D_{R_2})$: the left (right) detector of the second measurement branch. $D_{L_{N-1}}(D_{R_{N-1}})$: the left (right) detector of the $(N-1)$-th measurement branch. BS: Beam Splitter. FC: Fiber Channel.} \label{PMQCC_setup}
\end{figure}

Recently, twin-field quantum key distribution (TF-QKD) and phase-matching quantum key distribution (PM-QKD)~\cite{lucamarini2018overcoming,ma2018phase,tamaki2018information,wang2018twin,Cui2019Twin-Field,lin2018simple,yu2019sending,curty2019simple,maeda2019repeaterless,grasselli2019practical} are reported to overcome the repeater-less rate-distance limit~\cite{pirandola2017fundamental} of quantum key distribution (QKD). By introducing single-photon interference, these protocols achieve key generation rates scaling with the square-root of the channel transmittance $O(\sqrt{\eta})$ which exceeds the rate-distance limit without quantum repeaters. Here $\eta$ is the transmission rate of the optical channel between two users. At the same time, their measurement device can be controlled by an untrusted third party, which is also immune to all detector side-channel attacks~\cite{lo2012measurement}. These new types of QKD protocols have also been realized~\cite{wang2019beating,minder2019experimental,liu2019experimental,zhong2019proof,fang2020implementation} and shown to extend the distance of repeaterless fibre QKD to over 500 km~\cite{fang2020implementation}.

In this paper, we present a new QCC network protocol by combining ideas of phase-matching weak coherent pulses (WCPs) interference and post-selecting GHZ states, named as phase-matching quantum cryptographic conferencing (PM-QCC). As shown in Appendix~\ref{entanglement_distillation} and \ref{PMQCC}, successful WCPs interference events imply successful post-selection of multiparty GHZ states within the GHZ state basis:
\begin{equation}\begin{split}
  |\psi_{j,i_1i_2\cdots i_{N-1}}\rangle=&\frac{1}{\sqrt{2}}[|0i_1i_2\cdots i_{N-1}\rangle \\
                                          &+(-1)^j|1\bar{i}_1\bar{i}_2\cdots \bar{i}_{N-1}\rangle],
\end{split}
\end{equation}
where $j$ ($i_m$) $\in\{0,1\}$ is called phase (amplitude) bit, $1\leq m \leq N-1$, $\bar{i}_m$ is logical negation of $i_m$. Resorting to the entanglement distillation protocol~\cite{maneva2002improved}, one can distill the perfect N-qubit GHZ state:
\begin{equation}\label {cat_state}
  |\Phi^+\rangle=\frac{1}{\sqrt{2}}(|00\cdots 0\rangle + |11\cdots 1\rangle)_N,
\end{equation}
which can be used to generate secret key bits among $N$ parties.

In terms of the presented PM-QCC network, since the measurement device can be untrusted, it is immune to all detector side-channel attacks. Owing to its simpler setup structure compared with MDI-QCC networks based on GHZ analyzer~\cite{pan1998greenberger,qian2005universal}, one can extend PM-QCC to more users easily. Similar to the TF-QKD protocol, the key generation rate of the presented PM-QCC network can be improved to scale with $\eta^{N-1}$, whereas that of MDI-QCC network scales with $\eta^{N}$. Here, $\eta$ is the transmission rate of the optical channel from each party to the untrusted third party, Eve. Practically, there might be small-scale interference between $N'$ parties ($N'$ parties are near-neighbor connected, and $2\leq N' \leq N$) instead of perfect interference of $N$ parties. It is demonstrated that the small-scale $N'$-party PM-QCC can still be realized securely with key generation rates scaling with $\eta^{N'-1}$.

\section{PM-QCC Network}
Supposing that $N$ parties $P_1$, $P_2$, $\cdots$, $P_N$ plan to conduct a quantum cryptographic conference task, see Fig.~\ref{PMQCC_setup}. They can encode their random bits in their phase randomized coherent pulses. The encoded coherent pulses are sent to the untrusted third party, Eve, who is supposed to perform interference measurements. The $N$-party PM-QCC network works as follows:

\begin{enumerate}
  \item[Step.1] \textbf{Preparation}: Party $P_1$ randomly generates one bit $k_{1}\in\{0,1\}$ and one coherent pulse with a random phase $\phi_{1}\in[0,2\pi)$. Then, he encodes the random bit to the coherent pulse and get a phase randomized coherent pulses $|\mathrm{e}^{i(\phi_{1}+\pi k_{1})}\sqrt{\mu_1}\rangle$. Similarly, parties $P_2$, $\cdots$, $P_N$ prepare their phase randomized coherent pulses $|\mathrm{e}^{i(\phi_{2}+\pi k_{2})}\sqrt{\mu_{2}}\rangle$, $\cdots$, $|\mathrm{e}^{i(\phi_{N}+\pi k_{N})}\sqrt{\mu_{N}}\rangle$, respectively.

      As shown in Fig.~\ref{PMQCC_setup}, the settings for $P_1$, $P_N$ are different from that for $P_2$, $\cdots$, $P_{N-1}$ in the experimental setup. Thus, the intensities of the weak coherent pulses used by parties $P_1$, $P_N$ are set to be $\mu_1,\mu_N \in\{\frac{\mu}{2}>\frac{\nu}{2}>\frac{\omega}{2}>\frac{\tau}{2}>\cdots>0\}$, while the intensities for parties $P_2$, $\cdots$, $P_{N-1}$ are set to be $\mu_t \in\{\mu>\nu>\omega>\tau>\cdots>0\}$ ($2\leq t\leq N-1$). The pulses with intensity $\mu$ are used as signal pulses and the pulses with intensities $\{\nu,\omega,\tau,\cdots,0\}$ are used as decoy pulses.

  \item[Step.2] \textbf{Measurement}:  All the parties send their pulses directly to the untrusted third party Eve. By design, an honest Eve splits each pulse of parties $P_2$, $\cdots$, $P_{N-1}$ into two separated coherent pulses using $50:50$ beam splitters (BS1) to perform interference measurements as shown in Fig.~\ref{PMQCC_setup}. Eve measures the received pulses and records measurement results. Here, successful detection events are defined as coincidence clicks of $N-1$ measurement branches, within which only one detector clicks.

  \item[Step.3] \textbf{Announcement}: Eve announces measurement results for successful detection events. Then, all the parties announce their random phases $\phi_{1}$, $\phi_{2}$ $\cdots$ $\phi_{N}$ and their randomly chosen intensities $\mu_1,\mu_2,\cdots,\mu_N$, respectively.

  \item[Step.4] \textbf{Sifting}: When a successful detection event is announced by Eve, the $N$ parties $P_1$, $P_2$, $\cdots$, $P_N$ keep their random bits $k_{1}$, $k_{2}$ $\cdots$ $k_{N}$, respectively. A successful detection event is one of $2^{N-1}$ coincident click events in the set $\{D_{L_1}D_{L_2}\cdots D_{L_{N-1}}$, $ D_{R_1}D_{L_2}\cdots D_{L_{N-1}}$, $\cdots$, $D_{R_1}D_{R_2}\cdots D_{R_{N-1}}\}$. Here, $D_{L_l(R_l)}$ means that only the detector $D_{L_l}(D_{R_l})$ clicks in the $l$-th measurement branch. According to Eve's announcements, they cooperate to flip theirs bits to make their encoded phases the same with that of events $D_{L_1}D_{L_2}\cdots D_{L_{N-1}}$. Then, $P_1$, $P_2$,$\cdots$, $P_N$ keep their random bits only when the phase-matching conditions are satisfied: $|\phi_{1}-\phi_{2}|=0~\text{or}~\pi$, $|\phi_2-\phi_3|=0~\text{or}~\pi$, $\cdots$, $|\phi_{N-1}-\phi_N|=0~\text{or}~\pi$ and their intensities are $2\mu_1=\mu_t=2\mu_N$ ($2\leq t\leq N-1$). Then, according to their phase announcements, they cooperate to flip theirs kept random bits to be the same with that of $|\phi_{1}-\phi_{2}|=0$, $|\phi_2-\phi_3|=0$, $\cdots$, $|\phi_{N-1}-\phi_N|=0$ if they are not the case.
  \item[Step.5] \textbf{Parameter estimation and key distillation}: The above steps are repeated enough times to distill the raw key bits. From the data set generated by the signal pulses, the users can directly estimate the gain $Q_{\mu}$ and marginal quantum bit error rates (QBER) $E_{\mu,P_1P_2}^Z$, $E_{\mu,P_1P_3}^Z$, $\cdots$, $E_{\mu,P_1P_{N}}^Z$ from the measurement results. From the data set generated by the decoy pulses, the users can estimate the phase error $E^X_{\mu}$ according to decoy-state methods (see Appendix~\ref{decoy} for details). Finally, they distill private key bits by performing error correction and privacy amplification on the raw key.

  \end{enumerate}
For the coherent pulse interference measurement on the $l$-th ($1\leq l\leq N-1$) measurement branch, there would be only one detector click if the encoded phases of two pulses with equal intensities are matched, i.e. $D_{L_l}$(or $D_{R_l}$) would click if $\Delta\phi_l=|\phi_{l}+\pi k_{l}-(\phi_{l+1}+\pi k_{l+1})|=0$ (or $\pi$). This is vital in the security of PM-QCC network.

\begin{figure}[htbp]
  \centering
  \subfigure[]{
  \label{three_EDP_1}
  \includegraphics[width=0.35\textwidth]{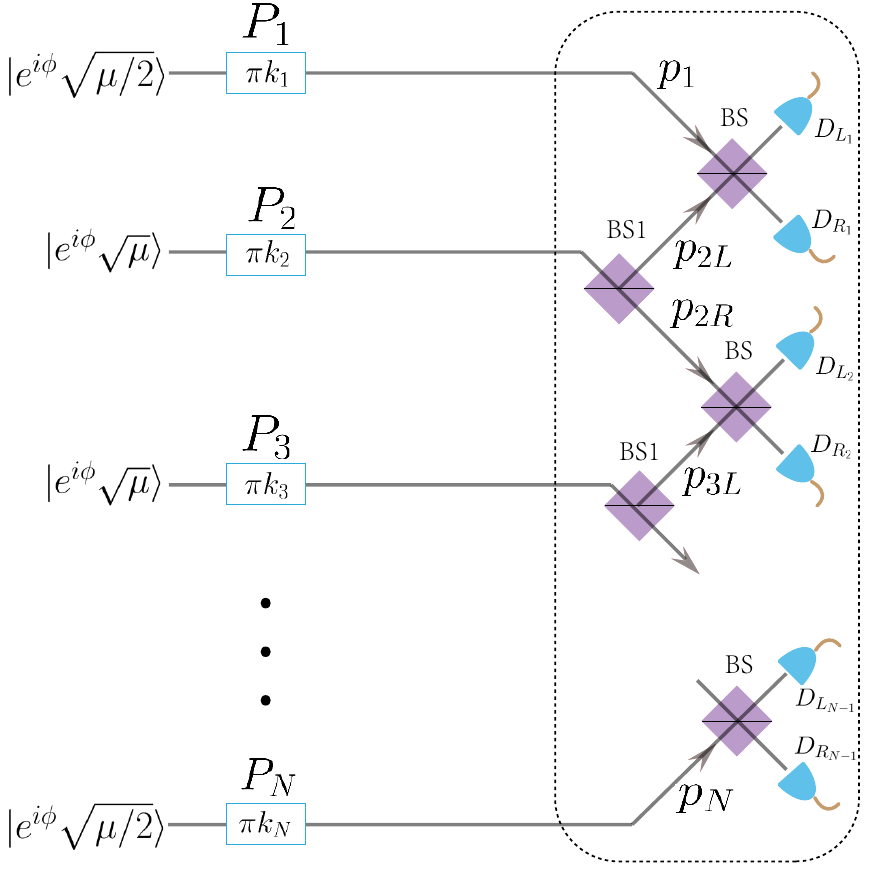}
  }
  \hspace{1in}
  \subfigure[]{
  \label{three_EDP_2}
  \includegraphics[width=0.35\textwidth]{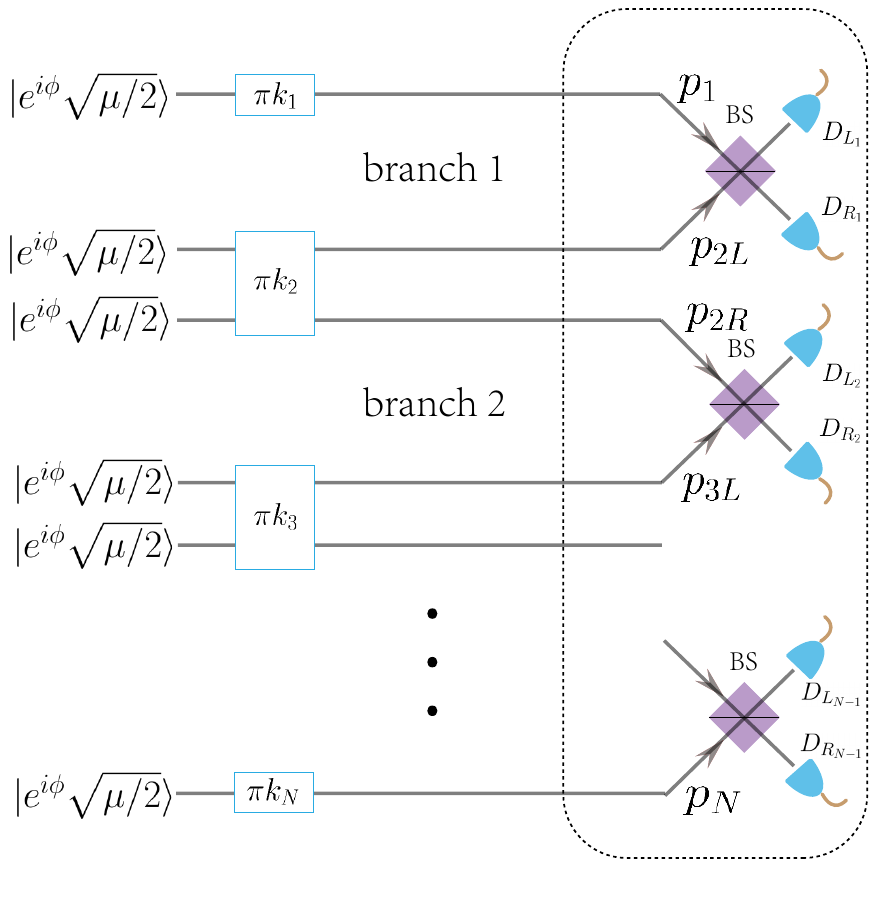}
  }
  \caption{(a) The PM-QCC protocol with phase matching condition $\phi_1=\phi_2=\cdots\phi_N=\phi$ satisfied. (b) Equivalent PM-QCC protocol after Eve's splitting with phase matching condition $\phi_1=\phi_2=\cdots\phi_N=\phi$ satisfied. Eve splits each pulse of parties $P_2$, $\cdots$, $P_{N-1}$ into two separated coherent pulses using a beam splitter BS1 to perform interference measurements. Once a success detection event is achieved, the encoded phases of $N$ parties are correlated with each other. $p_1,p_{2L},\cdots, p_N$: path modes after Eve's splitting. $D_{L_1}(D_{R_1})$: the left (right) detector of the first measurement branch. $D_{L_2}(D_{R_2})$: the left (right) detector of the second measurement branch. $D_{L_{N-1}}(D_{R_{N-1}})$: the left (right) detector of the $(N-1)$-th measurement branch. BS: Beam Splitter.}\label{EDP}
\end{figure}

In the above $N$-party PM-QCC network, the random phases $\phi_1$, $\phi_2$,$\cdots$, $\phi_N$ that $P_1$, $P_2$,$\cdots$, $P_N$ attach to their pulses are continuous. Thus, the precise phase-matching condition $|\phi_{m}-\phi_{m+1}|=0~\text{or}~\pi$ is hard to realize. Moreover, we suppose that the lasers of $P_1$, $P_2$,$\cdots$, $P_N$ are perfectly locked which is also impractical in experiments. To overcome these problems, we introduce the phase-compensation method~\cite{lucamarini2018overcoming,ma2018phase} that can help to conduct phase matching and phase reference. For an arbitrary party $P_m$, the phase interval $[0,2\pi)$ is cut into $M$ slices $\{\Delta_{j_m}\}$ with $0\leq j_m \leq M-1$, $\Delta_{j_m}=[\frac{2\pi}{M}j_m,\frac{2\pi}{M}(j_m+1))$. In the \textbf{Announcement} step, what $N$ parties $P_1$, $P_2$, $\cdots$, $P_N$ announce are their phase slice indexes $j_1$, $j_2$, $\cdots$, $j_N$ instead of their exact phases $\phi_1$, $\phi_2$, $\cdots$, $\phi_N$, respectively. Then, in the \textbf{Sifting} step, party $P_m$ and $P_{m+1}$ only need to compare their slices indexes, $|j_m+j_m^a-j_{m+1}| \mod M=0$ or $M/2$, where $j_m^a \in\{0,1,\cdots, M-1\}$ is an adjusted slice index to compensate the deviation of phase reference for parties $P_m$ and $P_{m+1}$. In practice, $j_m^a$ can be determined in the \textbf{Parameter estimation} step by minimizing the QBER. Although there will be intrinsic misalignment errors in the sifting induced by the coarse split of the phase interval, this makes phase-sifting practical without affecting the security~\cite{ma2018phase}.

\begin{figure}[htbp]
  \centering
  \includegraphics[width=0.45\textwidth]{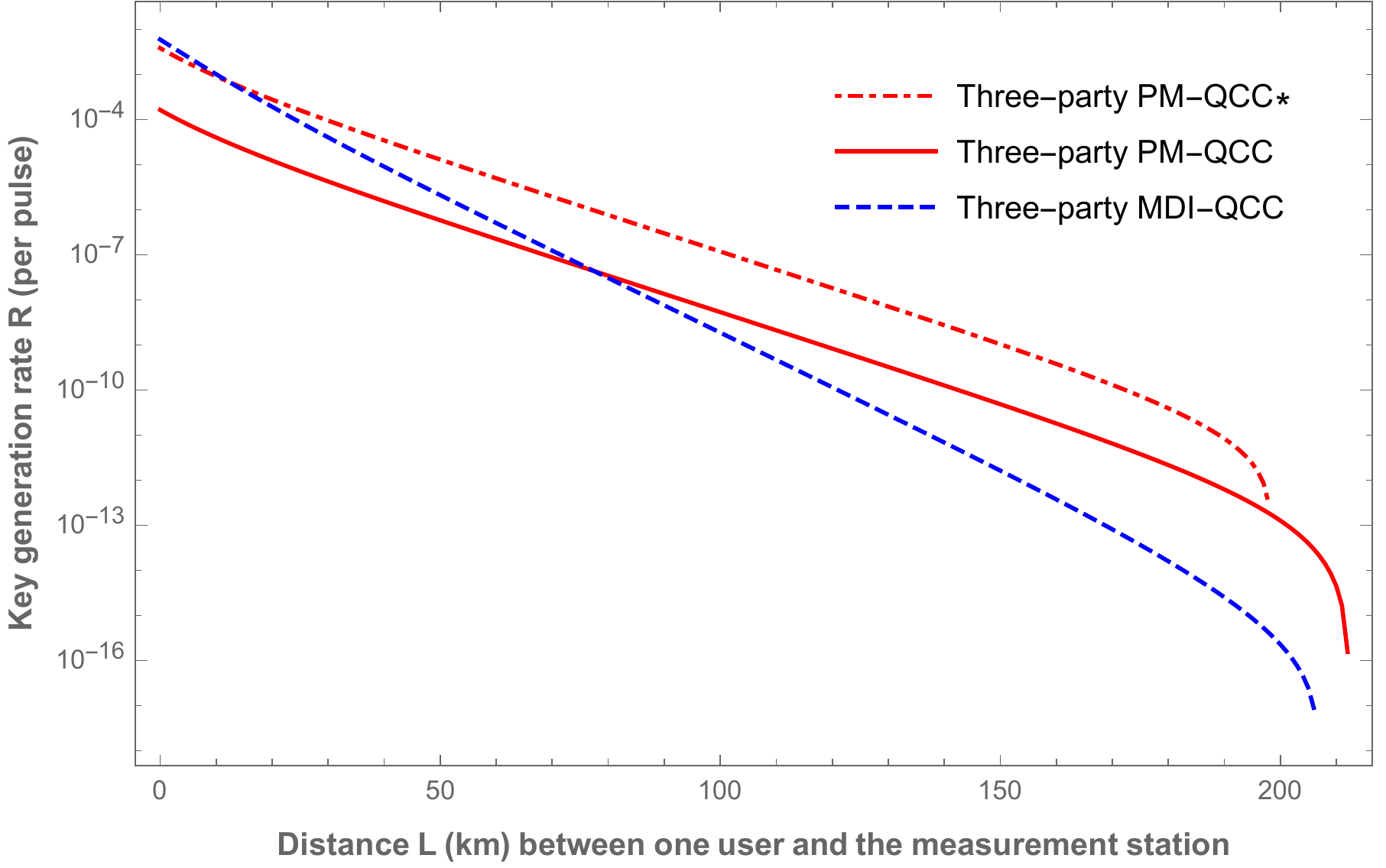}
  \caption{Key generation rate $R$ of three-party PM-QCC, three-party PM-QCC$*$ (without phase post-selection in signal pulses) and MDI-QCC network versus transmission distance $L$. The simulation result is obtained with parameters from Ref.~\cite{fu2015long} that the dark-count rate $p_d=1\times10^{-7}$, the loss rate of the channel $\alpha=0.2\mathrm{dB}/\mathrm{km}$, the detection efficiency $\eta_d=93\%$, the error correction efficiency $f=1.16$, the misalignment error for MDI-QCC $e_d^{\text{MDI}} = 1.5\%$ and the phase misalignment error for PM-QCC* $e_{\delta} = 1.5\%$. The number of phase slices $M$ for PM-QCC is optimized at different transmission distances.}\label{Three-party-keyRate}
\end{figure}

\section{Security Analysis}
Without loss of generality, we consider an entanglement-based protocol that party $P_m$ prepares entanglement states between his virtual qubits and his WCPs instead of directly preparing WCPs (see Appendix~\ref{PMQCC} for detail). Thus, its security analysis applies to the entanglement distillation argument~\cite{lo1999unconditional,shor2000simple,chen2005conference}. Following the entanglement distillation argument~\cite{lo1999unconditional,shor2000simple}, to generate a sequence of almost perfect secure key bits, parties $P_1$, $P_2$, $\cdots$, $P_N$ only need to share a sequence of almost perfect GHZ states in term of monogamy of entanglement~\cite{terhal2004entanglement,Koashi2004Monogamy}. Therefore, what we are facing now is to distill almost perfect GHZ states~\cite{maneva2002improved}.
%As shown in Fig.~\ref{EDP}, we illustrate the $N$ parties GHZ states distillation protocol.

As described in Step.4, when phase matching condition is satisfied, encoded random bits are kept. Without loss of generality, the phases are supposed to be $\phi_1=\phi_2=\cdots=\phi$. In Fig.~\ref{three_EDP_1}, the WCP with random phase $\phi$ of party $P_2$ arriving at the $50:50$ beam splitter BS1 is split into two WCPs with the same encoded phases.
\begin{equation}
      |\mathrm{e}^{i(\phi+\pi k_2)}\sqrt{\mu}\rangle \xrightarrow{\text{BS1}}|\mathrm{e}^{i(\phi+\pi k_2)}\sqrt{\mu/2}\rangle|\mathrm{e}^{i(\phi+\pi k_2)}\sqrt{\mu/2}\rangle,
\end{equation}
where $\phi+\pi k_2$ is the encoded phase of party $P_2$. The WCPs from party $P_2$ is split into two branches to interfere with $P_1$ and $P_3$ respectively. Similarly, WCPs from parties $P_3$, $\cdots$, $P_{N-1}$ are split. The third party, Eve, performs interfere measurement for all $N$ parties. Now, the protocol is equivalent to that of Fig.~\ref{three_EDP_2}.

Let us consider the entanglement based protocol of PM-QCC (see Appendix~\ref{PMQCC}). Once there is a success detection event, virtual qubits in $N$ parties are entangled together. After the distillation protocol (see Appendix~\ref{entanglement_distillation}), perfect GHZ states are shared between  $N$ parties. Finally, they can generate secret key bits from the distillation of the GHZ state~\cite{chen2007multi,chen2005conference,fu2015long}. The corresponding key generation rate is
  \begin{equation}\label{keyrate}
  \begin{split}
    R_{N-\text{party}}=& (\frac{2}{M})^{N-1} Q_{\mu}[1-f\cdot\max \{H(E_{\mu,P_1P_2}^Z), \\
                       &H(E_{\mu,P_1P_3}^Z), \cdots, H(E_{\mu,P_1P_{N}}^Z)\}-H(E_{\mu}^X)],
  \end{split}
  \end{equation}
where $H(x)=-x\log_2(x)-(1-x)\log_2(1-x)$ is the binary entropy function. The $E_{\mu,P_1P_{m}}^Z$ ($2 \leq m\leq N $) is the marginal QBER of parties $P_1$ and $P_{ m}$ and can be estimated from Eve's measurement results directly. The $E_{\mu}^X$ is the phase error rate which is an intrinsic error of the protocol and can be estimated with the help of the decoy-state method in experiments (see Appendix~\ref{Parameter} and \ref{decoy}). The $Q_{\mu}$ is the overall gain, and $\frac{2}{M}$ is induced by phase post-selection in the phase compensation method which can be optimized according to the experimental parameters~\cite{lucamarini2018overcoming,ma2018phase}.

As shown in Eq.~\ref{keyrate}, there is a prefactor $(\frac{2}{M})^{N-1}$ which is induced by the phase post-selecting process in the key generation rate. It might cause descending in key generation rate when the number of user increases. According to Appendix~\ref{PMQCC}, the PM-QCC protocol is still secure even when the phase choices in the signal pulses are announced before Eve's measurement if one can estimate the phase error accurately. Thus, the phase compensation method just provides a practical and secure way to align the phases for signal pulses. Then, if one can realize accurate and secure phase reference in his (or her) lab, the PM-QCC protocol can be improved to a version PM-QCC* without phase post-selection in signal pulses (see Appendix~\ref{without_PhaseSlice} for detail). It has also been demonstrated in new variants for TF-QKD and PM-QKD protocols~\cite{wang2018twin,curty2019simple,Cui2019Twin-Field,maeda2019repeaterless,grasselli2019practical}. In the PM-QCC*, the factor $(\frac{2}{M})^{N-1}$ can be improved to $1$, and the key generation rate is
\begin{equation}\label{}
  \begin{split}
    R_{N-\text{party}}^*=&Q_{\mu}^*[1-f\cdot\max \{H(E_{\mu,P_1P_2}^{Z*}), \\
                       &H(E_{\mu,P_1P_3}^{Z*}), \cdots, H(E_{\mu,P_1P_{N}}^{Z*})\}-H(E_{\mu}^{X*})].
  \end{split}
  \end{equation}
where $Q_{\mu}^*$ is the overall gain, $E_{\mu,P_1P_2}^{Z*}$, $E_{\mu,P_1P_2}^{Z*}$, $E_{\mu,P_1P_2}^{Z*}$ are marginal QBERs and $E_{\mu}^{X*}$ is the phase error rate. Need to note that the signal pulses from the parties can no longer be regarded as photon number states since the phase randomization for signal pulses has been cancelled out in the PM-QCC* protocol. Thus, the above mentioned decoy states discussion for the PM-QCC protocol becomes unsuitable for the PM-QCC* protocol, and more delicate decoy-state method is required to evaluate the phase error rate in the signals (see Appendix~\ref{without_PhaseSlice})~\cite{lin2018simple,curty2019simple,maeda2019repeaterless}. For example, as in~\cite{curty2019simple}, the estimation of phase error rate is converted to the estimation of the yields for the photon number state, which can be estimated using phase randomized decoy states. Thus, the phase randomized decoy states with different intensities can in principle be used to constraint the phase error rate $E_{\mu}^X$ tightly and we leave it for further studies.

\begin{figure}[htbp]
  \centering
  \includegraphics[width=0.45\textwidth]{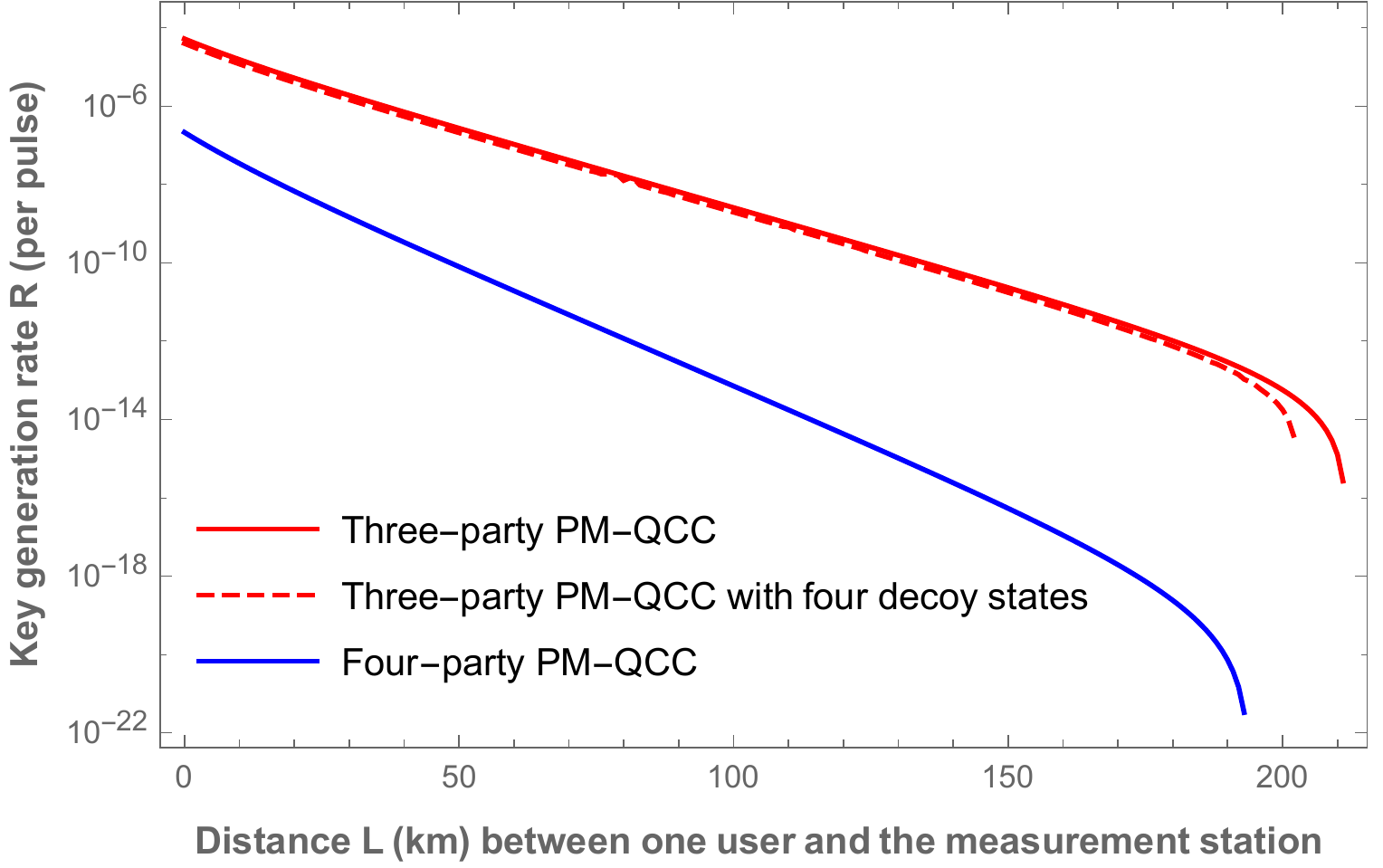}
  \caption{Key generation rate $R$ for $3$-party PM-QCC, $3$-party PM-QCC with four decoy states ($\nu$,$\omega$,$\tau$,0) and $4$-party PM-QCC  versus transmission distance $L$. Parameters adopted in simulation are derived from Ref.~\cite{yin2016measurement}: the dark-count rate $p_d=7.2\times 10^{-8}$, the loss rate of the channel $\alpha=0.2\mathrm{dB}/\mathrm{km}$, the detection efficiency $\eta_d=65\%$, the error correction efficiency $f=1.16$.}\label{N-party-keyRate}
\end{figure}

\section{Performance of PM-QCC network}
Without loss of generality, the channels between party $P_m$ and measurement station are supposed to be symmetric. To show the performance of PM-QCC network, we consider the 3-party PM-QCC network, 3-party PM-QCC* network and compare our protocol with MDI-QCC network~\cite{fu2015long} using the following experimental parameters: the intrinsic fiber channel loss $\alpha=0.2~\mathrm{dB}/\mathrm{km}$, detection efficiency of threshold single-photon detector $\eta_d=93\%$, dark-count rate $p_d=10^{-7}$, error correction efficiency $f=1.16$, misalignment error for MDI-QCC $e_d^{\text{MDI}} = 1.5\%$ and phase error for PM-QCC* $e_{\delta}^{\text{PM-QCC*}} = 1.5\%$. As shown in Fig.~\ref{Three-party-keyRate}, one can see that the key generation rates of PM-QCC can be well beyond that of MDI-QCC around $L =80$ km. The key rate is improved by approximately 2 orders of magnitude around $L=200$ km. For the PM-QCC* without phase-matching condition in signal pulses, the key generation can be well beyond that MDI-QCC around $L=12$ km. This improvement mainly comes from the fact that the key generation rate of the presented PM-QCC network can be improved to scale with $\eta^{N-1}$, whereas that of MDI-QCC network scales with $\eta^{N}$ (see Appendix~\ref{comparison}).
\begin{table}[h]
  \caption{The performance for PM-QCC network at $N=3$. The key generation rate $R$, mean photon number $\mu$ and phase slice number $M$ are optimized with $p_d=7.2\times10^{-8}$, $\eta_d=65\%$, $f=1.16$ and $\alpha=0.2~\mathrm{dB}/\mathrm{km}$ at different transmission distance $L$.}\label{table:simulation}
  \begin{tabular*}{0.45\textwidth}{@{\extracolsep{\fill}} c c c c  }
  \hline\hline
  $R~(bits~per~pulse)$ &$L~(\text{Km})$ & $\mu$ & $M$ \\
  \hline
  $2.6989\times10^{-7}$ & $50$ & $0.1333$ & $13$ \\
  $1.6227\times10^{-8}$ & $80$ & $0.1299$ & $13$\\
  $2.5332\times10^{-9}$ & $100$ & $0.1291$ & $13$\\
  $2.2928\times10^{-11}$ & $150$ & $0.1263$ & $13$\\
  $2.6206\times10^{-14}$ & $200$ & $0.1239$ & $17$\\
  \hline\hline
 \end{tabular*}
\end{table}

To demonstrate the scalability and the decoy-state method of PM-QCC network, we simulate the PM-QCC network at $N=3$ parties with infinite decoy states and four decoy states ($\nu>\omega>\tau> 0$). With experimental parameters given in~\cite{yin2016measurement} that detection efficiency $\eta_d=65\%$, dark-count rate $p_d=7.2\times10^{-8}$, the performance of PM-QCC network at $N=3$ parties are presented in Fig.~\ref{N-party-keyRate}. The longest transmission distance between one user and the measurement station of PM-QCC is beyond 200 km at $N=3$ parties. Remarkably, the longest transmission distance between two users is over 400 km by special arrangement. The optimized weak coherent states $\mu$ and phase slice numbers $M$  for given parameters at $N=3$ are presented in Table.~\ref{table:simulation}. Meanwhile, the key generation rate for $3$-party PM-QCC with four decoy states is optimized over $\mu$, $\nu$, $\omega$ and $\tau$ for given parameters and $M=13$. For example, the key generation rate is $R=1.7327\times 10^{-11}$ (bits per pulse) at $L=150$ (km) with $\mu=0.104815$, $\nu=0.0204583$, $\omega=0.0182017$, $\tau=9.27216\times10^{-5}$. Furthermore, we simulate the PM-QCC at $N=4$ parties, and the simulation results are present in Fig.~\ref{N-party-keyRate}.

According to the above discussion, it is feasible to realize the PM-QCC network for three and even more parties with current experimental technology. Meanwhile, the PM-QCC* without phase post-selection in signal pulses (see Appendix~\ref{without_PhaseSlice}) can be realized with further optimization and accurate phase reference in a long distance.

Practically, there might be interferences of only $N'$ parties ($N'$ parties are near-neighbor connected, and $2\leq N' \leq N$) instead of perfect interference of $N$ parties. In this case, according to Step.1 of the PM-QCC network, the weak coherent pulses prepared by parties at broken points are $|\sqrt{\mu}\rangle$ instead of $|\sqrt{\mu/2}\rangle$ (the encoded phase is omit here). While, as shown in Fig.~\ref{PMQCC_setup}, the intensities of weak coherent pulses arriving at the third party are equal in an inference branch. Therefore, higher amounts of weak coherent pulses are lost during the transmission for broken points compared with that for unbroken points. It is demonstrated that the secure reduced small-scale PM-QCC networks can also be constructed among $N'$ parties with key generation rate $R_{\text{reduced} PM-QCC}\varpropto \eta^{N'-1}$ (see Appendix~\ref{small} for detail).

\section{Conclusion and Outlook}
Based on the multiparty weak coherent pulses interference, we present a new protocol named as phase matching quantum cryptographic conferencing (PM-QCC) network that can distribute information-theoretic secure keys among $N$ parties. In the merit of simpler setup, the PM-QCC network can be conveniently generalized to $N$ parties and can go beyond the existing QCC networks. Firstly, similarly to the MDI-QCC network, the PM-QCC network is immune to all detector side-channel attacks since the measurement device can be untrusted. Secondly, compared with the MDI-QCC networks based on the GHZ analyzer, the PM-QCC can be more easily extended to multiple users due to simpler setup structure. Thirdly, the key generation rate of the presented PM-QCC network can be improved to scale with $\eta^{N-1}$, whereas that of MDI-QCC network scales with $\eta^{N}$. Fourthly, considering practical cases that small-scale interferences between $N'$ parties instead of perfect interferences of $N$ parties, the small-scale $N'$-party PM-QCC can still be realized. Finally, based on the setup of the PM-QCC network, GHZ state distribution networks can be constructed directly, which may be of great potential for other implementations in quantum information science.

During the preparation of this manuscript, a related work based on the post-selection of W state~\cite{dur2000three} has been reported in Ref.~\cite{grasselli2019conference} which is named as conference key agreement with single-photon interference (single-photon CKA). Compared with the single-photon CKA, the proposed protocol is an essentially different protocol. Specifically, the proposed protocol is a MDI prepare-and-measure scheme, while, the single-photon CKA cannot be cast to a MDI prepare-and-measure scheme in which all the parties have to measure their local qubits and trust the measurement results. Meanwhile, the signal qubits sent to the measurement station cannot be replaced by coherent states, and the local qubits have to resort to quantum memories in the single-photon CKA. Thus, the feasibility of the single-photon CKA requires further investigation~\cite{grasselli2019conference}.

\section{Acknowledgments}
We acknowledge Feihu Xu for insightful discussion. This work has been supported by the Chinese Academy of Science, the National Fundamental Research Program, the National Natural Science Foundation of China (Grants No.11575174, No.11374287, No.11574297, No.11875173, and No.11674193), the National Key R$\&$D Program of China (Grants No.2017YFA0303900 and No.2017YFA0304004), the Anhui Initiative in Quantum Information Technologies, as well as the Zhongguancun Haihua Institute for Frontier Information Technology.

\begin{appendix}
\section{Distillation of GHZ State}\label{entanglement_distillation}
Inspired by the quantum key distribution protocol based on entanglement distillation~\cite{lo1999unconditional,shor2000simple}, multi-party quantum conference key distribution protocols based on entanglement distillation are proposed to securely distribute random bits between multiple users~\cite{chen2007multi,chen2005conference,fu2015long}. The security of phase matching quantum cryptographic conferencing network (PM-QCC) is based on the distillation of N-qubit GHZ state~\cite{maneva2002improved}
\begin{equation}\label {cat_state}
  |\Phi^+\rangle=\frac{1}{\sqrt{2}}(|00\cdots 0\rangle + |11\cdots 1\rangle)_N,
\end{equation}
which is stabilized by a group of stabilizer generators,
\begin{equation}
  \begin{split}
    &S_0=XXXX\cdots X,\\
    &S_1=ZZII\cdots I,\\
    &S_2=ZIZI\cdots I,\\
    &S_3=ZIIZ\cdots I,\\
    &\vdots\\
    &S_{N-1}=ZIII\cdots Z,
  \end{split}
\end{equation}
where $X=\left(
           \begin{array}{cc}
             0 & 1 \\
             1 & 0 \\
           \end{array}
         \right)
$, $Z=\left(
           \begin{array}{cc}
             1 & 0 \\
             0 & -1 \\
           \end{array}
         \right)
$, $I=\left(
           \begin{array}{cc}
             1 & 0 \\
             0 & 1 \\
           \end{array}
         \right)
$ are Pauli matrices. The corresponding N-qubit GHZ state basis is
\begin{equation}\begin{split}
  |\psi_{j,i_1i_2\cdots i_{N-1}}\rangle=&\frac{1}{\sqrt{2}}[|0i_1i_2\cdots i_{N-1}\rangle \\
                                          &+(-1)^{j}|1\bar{i}_1\bar{i}_2\cdots \bar{i}_{N-1}\rangle],
\end{split}
\end{equation}
where $j, i_m \in\{0,1\}$, $1\leq m \leq N-1$, $\bar{i}_m$ is logical negation of $i_m$. If $j=1$ ($i_m=1$), the basis vector is the $-1$ eigenvalue of $S_0$ ($S_m$). It means that there is a phase error (bit error) to the original GHZ state. Thus, $j$ (or $i_m$) is also called phase (or amplitude) bit. Using the multipartite hashing method~\cite{maneva2002improved}, the yield of distillation of the pure N-qubit GHZ state is
\begin{equation}\label{distillation}
\begin{split}
  D=&1-\max\{H(E_{\mu,P_1P_2}^Z),H(E_{\mu,P_1P_3}^Z),\\
      &\cdots, H(E_{\mu,P_1P_N}^Z)\}-H(E_{\mu}^X),
\end{split}
\end{equation}
where $E_{\mu,P_1P_m}^Z$ represents the bit flip error rate of parties $P_1$ and $P_m$ corresponding to the stabilizer $S_{m-1}$. The $E_{\mu}^X$ is the phase flip error corresponding to the stabilizer $S_0$, $H(x)=-x\log_2(x)-(1-x)\log_2(1-x)$ is the binary entropy function.
\begin{figure}[]
  \centering
  \subfigure[]{
  \label{PM_QCC_EDP}
  \includegraphics[width=0.3\textwidth]{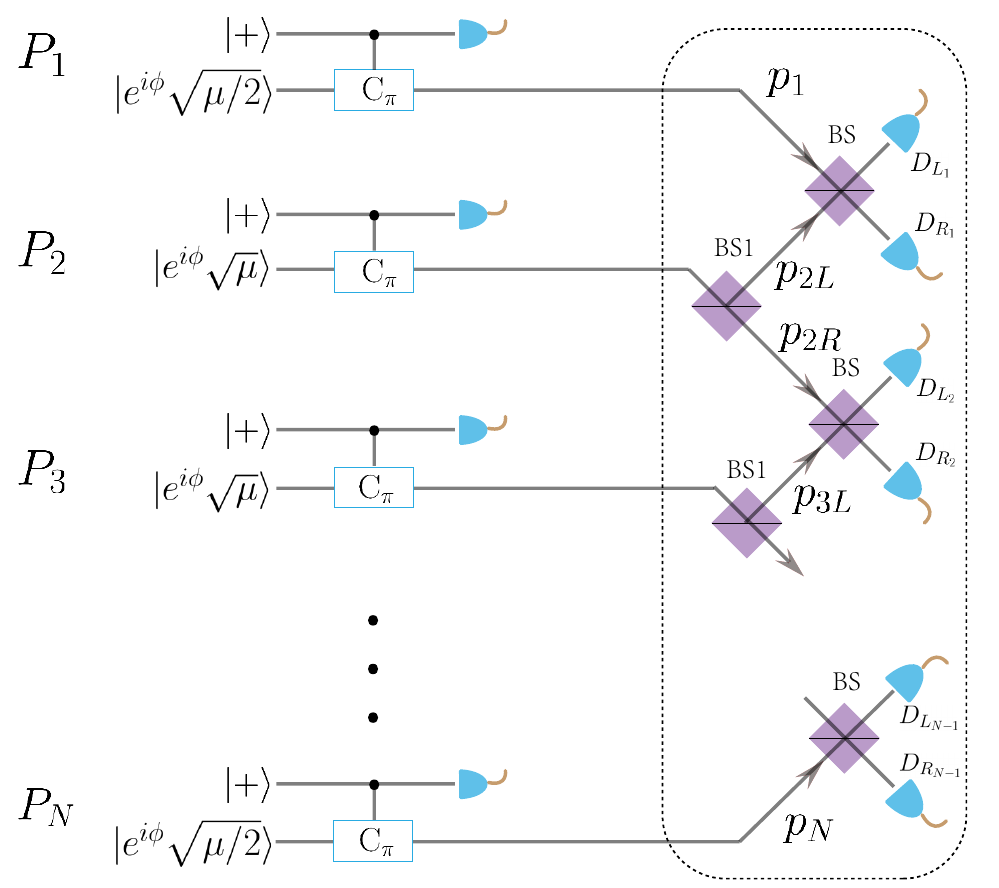}
  }
  \hspace{0.5in}
  \subfigure[]{
  \label{three_EDP_3}
  \includegraphics[width=0.3\textwidth]{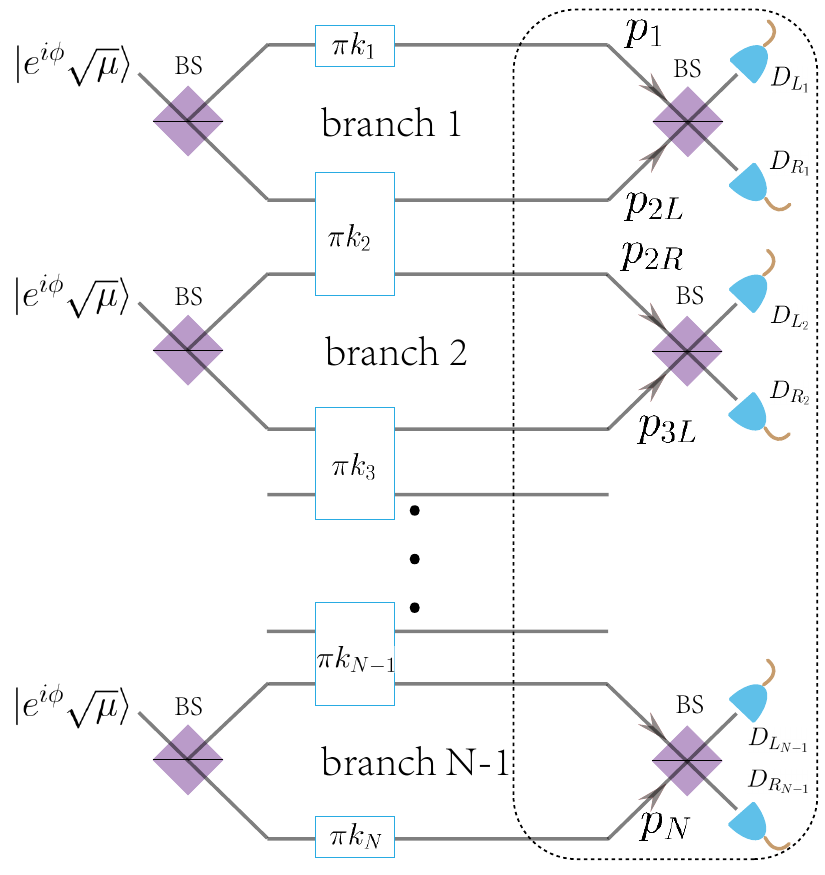}
  }
  \caption{(a) The entanglement based version of $N$-party PM-QCC network. (b) The equivalent entanglement based version of $N$-party PM-QCC network with virtual sources after Eve's splitting. In the Fig.~\ref{PM_QCC_EDP} and Fig.~\ref{three_EDP_3}, random phases are supposed to be $\phi_1=\phi_2=\cdots=\phi$. Here, we omit the virtual qubits for simplicity in subfigure (b). $p_1,p_{2L},\cdots, p_N$: path modes after Eve's splitting. $D_{L_1}(D_{R_1})$: the left (right) detector of the first measurement branch. $D_{L_2}(D_{R_2})$: the left (right) detector of the second measurement branch. $D_{L_{N-1}}(D_{R_{N-1}})$: the left (right) detector of the $(N-1)$-th measurement branch. BS: Beam Splitter. $\text{C}_{\pi}$: the control phase gate.}
  \label{evolution}
\end{figure}
%\begin{figure*}[]
%  \centering
%  \subfigure[]{
%  \label{PM_QCC_EDP}
%  \includegraphics[width=0.23\textwidth]{PM_QCC_EDP}
%  }
%  \hspace{0.5in}
%  \subfigure[]{
%  \label{three_EDP_3}
%  \includegraphics[width=0.2\textwidth]{three_EDP_3}
%  }
%  \caption{(a) The entanglement based version of $N$-party PM-QCC network. (b) The equivalent entanglement based version of $N$-party PM-QCC network with virtual sources after Eve's splitting. In the Fig.~\ref{PM_QCC_EDP} and Fig.~\ref{three_EDP_3}, random phases are supposed to be $\phi_1=\phi_2=\cdots=\phi$. Here, we omit the virtual qubits for simplicity in subfigure (b). $p_1,p_{2L},\cdots, p_N$: path modes after Eve's splitting. $D_{L_1}(D_{R_1})$: the left (right) detector of the first measurement branch. $D_{L_2}(D_{R_2})$: the left (right) detector of the second measurement branch. $D_{L_{N-1}}(D_{R_{N-1}})$: the left (right) detector of the $(N-1)$-th measurement branch. BS: Beam Splitter. $\text{C}_{\pi}$: the control phase gate.}
%  \label{evolution}
%\end{figure*}
\section{Security Analysis for PM-QCC}\label{PMQCC}
Without loss of generality, we consider an entanglement based version that party $P_m$ ($1\leq m \leq N$) prepares entanglement states between virtual qubits and his WCPs instead of directly preparing WCPs.  Thus, its security analysis applies to the entanglement distillation argument~\cite{lo1999unconditional,shor2000simple,chen2005conference}. As shown in Fig.~\ref{PM_QCC_EDP}, there is a virtual qubit at each party $$|+\rangle=\frac{1}{\sqrt{2}}(|0\rangle +|1\rangle).$$ The party $P_m$ prepares an entanglement state using a control phase gate $C_{\pi}=|0\rangle\langle 0|U_{o}+|1\rangle\langle 1|U_{\pi}$ between the virtual and weak coherent pulse(WCP) that
\begin{equation}
  |\Psi\rangle_{m}=\frac{1}{\sqrt{2}}[|0\rangle|\mathrm{e}^{i\phi}\sqrt{\mu_m}\rangle+|1\rangle|\mathrm{e}^{i(\phi+\pi)}\sqrt{\mu_m}\rangle],
\end{equation}
where $U_{0(\pi)}$ will attach a phase of $0$ ($\pi$) to the WCP. Without loss of generality, the phases of parties $P_1$, $P_2$, $\cdots$, $P_N$ are supposed to be $\phi_1=\phi_2=\cdots=\phi_N=\phi$. The WCPs are sent to untrusted third party, Eve, to perform interference measurements with other parties.  While, the virtual qubits are kept at each party. As is stated in the main text, the WCP with random phase $\phi$ of party $P_m$ passing through the $50:50$ beam splitter BS1 is split into two WCPs with the same encoded phases.
\begin{equation}
      |\mathrm{e}^{i(\phi+\pi k_m)}\sqrt{\mu}\rangle\xrightarrow{\text{BS1}}|\mathrm{e}^{i(\phi+\pi k_m)}\sqrt{\mu/2}\rangle\otimes|\mathrm{e}^{i(\phi+\pi k_m)}\sqrt{\mu/2}\rangle,
\end{equation}
where $\phi+\pi k_m$ is the encoded phase of party $P_m$. The protocol is the equivalent entanglement based protocol of Fig.~\ref{three_EDP_2}. Since the neighbor WCPs are of the same intensity, the only difference is their phases. Thus, the WCPs in each branch can be regarded as from one virtual WCP source $|\mathrm{e}^{i\phi}\sqrt{\mu}\rangle$, and the protocol is straightforwardly equivalent to that in Fig.~\ref{three_EDP_3}. In the protocol of Fig.~\ref{three_EDP_3}, the $N$-party sate evolves as
\begin{widetext}
\begin{equation}\label{bit_phase_expand}
    \begin{split}
       &|+\rangle_{P_1}|+\rangle_{P_2}\cdots|+\rangle_{P_{N}} \sum_{n_1=0}^{\infty}\mathrm{e}^{-\mu/2}\frac{(\mathrm{e}^{i\phi}\sqrt{\mu}C_1^{\dag})^{n_1}}{n_1!} \sum_{n_2=0}^{\infty}\mathrm{e}^{-\mu/2}\frac{(\mathrm{e}^{i\phi}\sqrt{\mu}C_2^{\dag})^{n_2}}{n_2!}\cdots \sum_{n_{N-1}=0}^{\infty}\mathrm{e}^{-\mu/2}\frac{(\mathrm{e}^{i\phi}\sqrt{\mu}C_{N-1}^{\dag})^{n_{N-1}}}{n_{N-1}!}|\text{vac}\rangle\\
       &\xrightarrow{\text{BS}}\sum_{n_1,n_2,\cdots,n_{N-1}=0}^{\infty}|+\rangle_{P_1}|+\rangle_{P_2}\cdots|+\rangle_{P_{N}}\frac{\mathrm{e}^{-\mu/2}(\mu)^{\frac{n_1}{2}}}{n_1!}(\frac{\mathrm{e}^{i\phi}p_1^{\dag}+\mathrm{e}^{i\phi}p_{2L}^{\dag}}{\sqrt{2}})^{n_1}\frac{\mathrm{e}^{-\mu/2}(\mu)^{\frac{n_2}{2}}}{n_2!}(\frac{\mathrm{e}^{i\phi}p_{2R}^{\dag}+\mathrm{e}^{i\phi}p_{3L}^{\dag}}{\sqrt{2}})^{n_2}
       \cdots\\
       &\frac{\mathrm{e}^{-\mu/2}(\mu)^{\frac{n_{N-1}}{2}}}{n_{N-1}!}(\frac{\mathrm{e}^{i\phi}p_{(N-1)R}^{\dag}+\mathrm{e}^{i\phi}p_{N}^{\dag}}{\sqrt{2}})^{n_{N-1}}|\text{vac}\rangle\\
       &\xrightarrow{C_{\pi}}\frac{1}{2^{N/2}}\sum_{n_1,n_2,\cdots,n_{N-1}=0}^{\infty}\frac{\mathrm{e}^{-(N-1)\mu/2}(\mu)^{\frac{n_1+n_2+\cdots+n_{N-1}}{2}}}{n_1!n_2!\cdots n_{N-1}!}[|0\rangle|0\rangle\cdots|0\rangle(\frac{\mathrm{e}^{i\phi}p_1^{\dag}+\mathrm{e}^{i\phi}p_{2L}^{\dag}}{\sqrt{2}})^{n_1}(\frac{\mathrm{e}^{i\phi}p_{2R}^{\dag}+\mathrm{e}^{i\phi}p_{3L}^{\dag}}{\sqrt{2}})^{n_2}
       \cdots\\
       &(\frac{\mathrm{e}^{i\phi}p_{(N-1)R}^{\dag}+\mathrm{e}^{i\phi}p_{N}^{\dag}}{\sqrt{2}})^{n_{N-1}}+ \cdots\\
       &+|1\rangle|1\rangle\cdots|1\rangle(\frac{-\mathrm{e}^{i\phi}p_1^{\dag}-\mathrm{e}^{i\phi}p_{2L}^{\dag}}{\sqrt{2}})^{n_1}(\frac{-\mathrm{e}^{i\phi}p_{2R}^{\dag}-\mathrm{e}^{i\phi}p_{3L}^{\dag}}{\sqrt{2}})^{n_2}
       \cdots (\frac{-\mathrm{e}^{i\phi}p_{(N-1)R}^{\dag}-\mathrm{e}^{i\phi}p_{N}^{\dag}}{\sqrt{2}})^{n_{N-1}}]|\text{vac}\rangle\\
       &=\frac{1}{2^{(N-1)/2}}\sum_{n_1,n_2,\cdots,n_{N-1}=0}^{\infty}\frac{\mathrm{e}^{-(N-1)\mu/2}(\mu)^{\frac{n_1+n_2+\cdots+n_{N-1}}{2}}}{n_1!n_2!\cdots n_{N-1}!}\\
       &\cdot\{\sum_{i_1,i_2,\cdots,i_{N-1}\in\{0,1\}}\frac{1}{\sqrt{2}}[|0 i_1 i_2\cdots i_{N-1}\rangle+(-1)^{n_1+n_2+\cdots+ n_{N-1}}|1 \bar{i}_1 \bar{i}_2\cdots \bar{i}_{N-1}\rangle]\\
       &[\frac{\mathrm{e}^{i\phi}p_1^{\dag}+(-1)^{i_1}\mathrm{e}^{i\phi}p_{2L}^{\dag}}{\sqrt{2}}]^{n_1}[\frac{(-1)^{i_1}\mathrm{e}^{i\phi}p_{2R}^{\dag}+(-1)^{i_2}\mathrm{e}^{i\phi}p_{3L}^{\dag}}{\sqrt{2}}]^{n_2}
       \cdots[\frac{(-1)^{i_{N-2}}\mathrm{e}^{i\phi}p_{(N-1)R}^{\dag}+(-1)^{i_{N-1}}\mathrm{e}^{i\phi}p_{N}^{\dag}}{\sqrt{2}}]^{n_{N-1}}\}|\text{vac}\rangle,\\
       &=\frac{1}{2^{(N-1)/2}}\cdot\sum_{i_1,i_2,\cdots,i_{N-1}\in\{0,1\}}\{\frac{1}{\sqrt{2}}[|0 i_1 i_2\cdots i_{N-1}\rangle+|1 \bar{i}_1 \bar{i}_2\cdots \bar{i}_{N-1}\rangle]\sqrt{p_{\text{even}}}|\text{even}\rangle_{\mu}\\
       &+\frac{1}{\sqrt{2}}[|0 i_1 i_2\cdots i_{N-1}\rangle-|1 \bar{i}_1 \bar{i}_2\cdots \bar{i}_{N-1}\rangle]\sqrt{p_{\text{odd}}}|\text{odd}\rangle_{\mu}\},
    \end{split}
\end{equation}
\end{widetext}
where $C_i^{\dag}$ is the creation operator of the $i$-th virtual source, $|\text{vac}\rangle$ is the vacuum state, $\sqrt{p_{\text{even}}}$ and $\sqrt{p_{\text{odd}}}$ are normalized coefficients of pure state $|\text{even}\rangle_{\mu}$ and $|\text{odd}\rangle_{\mu}$(see Eq.~\ref{coherentparity} to Eq.~\ref{podd}) , $p_1^{\dag},p_{2L}^{\dag},p_{2R}^{\dag},\cdots$ are the creation operator of the corresponding path mode after the BS1s. The BSs act as
  $$\text{BS}=\frac{1}{\sqrt{2}}\left(
                                 \begin{array}{cc}
                                   1 & 1 \\
                                   1 & -1 \\
                                 \end{array}
                               \right).$$
   For example, considering the beams interfere at the second BS shown in the first branch of Fig.~\ref{three_EDP_3}. For input beams optical modes $p_1^{\dag}$ and $p_{2L}^{\dag}$ , the output modes $L_1^{\dag} $ and $R_1^{\dag}$  are
  \begin{equation}\label{BS}
    \left(
       \begin{array}{c}
         L_1^{\dag} \\
         R_1^{\dag} \\
       \end{array}
     \right)=\frac{1}{\sqrt{2}}\left(
                                 \begin{array}{cc}
                                   1 & 1 \\
                                   1 & -1 \\
                                 \end{array}
                               \right)\cdot    \left(
       \begin{array}{c}
         p_1^{\dag} \\
         p_{2L}^{\dag} \\
       \end{array}
     \right),
  \end{equation}
where $L_1^{\dag}$($R_1^{\dag}$) means the output path mode to detector $D_{L_1}^{\dag}$($D_{R_1}^{\dag}$) for branch 1 in Fig.~\ref{three_EDP_3}. Then, we have
\begin{equation}
  \frac{p_1^{\dag}+(-1)^{i_1}p_{2L}^{\dag}}{\sqrt{2}}\xrightarrow{\text{BS}}
  \begin{cases}
    L_1^{\dag} & \text{if  }  i_1 =0,\\
    R_1^{\dag} & \text{if  }  i_1 =1.
  \end{cases}
\end{equation}

Thus, when $i_1=0$ only detector $D_{L_1}$ clicks, while when $i_1=1$ only detector $D_{R_1}$ clicks. From Eq.~\ref{bit_phase_expand}, once there is a success coincidence event that only one detector clicks in each branch, a GHZ state of $N$ parties is post-selected successfully such that
\begin{equation}
\begin{split}
   |\Psi_{j,i_1i_2\cdots i_{N-1}}\rangle &= \frac{1}{\sqrt{2}}[|0 i_1 i_2\cdots i_{N-1}\rangle\\
   &+(-1)^{j}|1 \bar{i}_1 \bar{i}_2\cdots \bar{i}_{N-1}\rangle],
\end{split}
\end{equation}
where, $j=n_1+n_2+\cdots+n_{N-1}$. One can obtain the phase error correlation immediately
 \begin{equation}\label{phase_N_party}
   e_{n_1,n_2,\cdots,n_{N-1}}^X= \left\{
      \begin{array}{ll}
        1, & \hbox{for $j\in\text{\textbf{odd}}$,} \\
        0, & \hbox{for $j\in\text{\textbf{even}}$.}
      \end{array}
    \right.
  \end{equation}
The phase error rate is determined by different photon number components. Note that similar phase-error property is shown in a improved analysis of PM-QKD~\cite{zeng2019symmetry}. When $n_1+n_2+\cdots+n_{N-1}\in\text{\textbf{odd}}$, $e_{n_1,n_2,\cdots,n_{N-1}}^X=1$ and when $n_1+n_2+\cdots+n_{N-1}\in\text{\textbf{even}}$, $e_{n_1,n_2,\cdots,n_{N-1}}^X=0$. Using the correlation of Eq.~\ref{phase_N_party} and Eq.~\ref{bit_phase_expand}, the total phase error rate $E^X_{\mu}$can be estimated as

\begin{equation}\label{phase_error}
\begin{split}
  E^X_{\mu}=&p_{\text{odd}}\cdot \frac{Y^{\text{odd}}_{\mu}}{Q_{\mu}},
\end{split}
\end{equation}
%\begin{equation}\label{phase_error}
%  E^X_{\mu}=\sum_{n_1,n_2,\cdots,n_{N-1}} q_{n_1,n_2,\cdots,n_{N-1}} e_{n_1,n_2,\cdots,n_{N-1}}^X,
%\end{equation}
where $Y^{\text{odd}}_{\mu}$ is the overall yield for odd number component, and $Q_{\mu}$ is the overall gain of signal pulses.

Following the entanglement distillation argument~\cite{lo1999unconditional,shor2000simple}, to generate a sequence of almost perfectly secure key bits, $P_1$, $P_2$, $\cdots$, $P_N$ only need to share a sequence of almost perfect GHZ states in term of monogamy of entanglement~\cite{terhal2004entanglement,Koashi2004Monogamy}. From Eq.~\ref{distillation}, the key generation rate of the PM-QCC network is
\begin{equation}\label{distillation_supp}
  \begin{split}
    R_{N-\text{party}}=&(\frac{2}{M})^{N-1}Q_{\mu}[1-f\cdot\max \{H(E_{\mu,P_1P_2}^Z), \\
                       &H(E_{\mu,P_1P_3}^Z), \cdots, H(E_{\mu,P_1P_{N}}^Z)\}-H(E_{\mu}^X)],
  \end{split}
  \end{equation}
where $(\frac{2}{M})^{N-1}$ is the prefactor induced by phase post-selection which can be optimized according to the experimental parameters~\cite{lucamarini2018overcoming,ma2018phase}.

What is counter-intuitive in the security analysis of the phase-matching protocol is that parties $P_1$, $P_2$, $\cdots$, $P_N$ will announce their random phases $\phi_1$, $\phi_2$, $\cdots$, $\phi_N$ after Eve's announcements of his measurement results. If the phases are not announced by $P_1$, $P_2$, $\cdots$, $P_N$, the weak coherent pulses can be regarded as the mixture of different photon number states in which their phases are meaningless to the untrusted party, Eve. While, after the announcement, their pulses can no longer be regarded as the mixture of photon number states. Here, we will show that the PM-QCC protocol is secure against the phase announcements.

Firstly, the random phases are announced after Eve's announcements. Thus, Eve's announcement strategies cannot depend on parties $P_1$, $P_2$, $\cdots$, $P_N$'s phase information.

Secondly, the security analysis is based on entanglement distillation. After the phase announcement, one can still distill perfect GHZ states which are decoupled from Eve according to the monogamy of entanglement. Specifically, let us consider the Beam splitting attack as an example in which Eve manages to get the key information using the announced phases.

In the Beam splitting attack, Eve can modulate transmission rates of the channels. For example, she using a beam splitter with transmission rate $\eta$ to simulate a lossy channel. The reflection signal beams are intercepted to Eve's registers, then the transmission beams are sent to interferometer through perfect channels. After the phase announcements, Eve would extract some information from the intercepted beams according to the phase announcements. From Eq.~\ref{bit_phase_expand}, we know that the state arriving at the detectors can be sort to $n_1+n_2+\cdots+n_{N-1}\in \text{\textbf{even}}$ or $\text{\textbf{odd}}$. Without loss of generality, we consider only the components that would result in the detection events $D_{L_1}D_{L_2}\cdots D_{L_{N-1}}$ in Eq.~\ref{bit_phase_expand}. The correlated components are
\begin{equation}\label{Lclikcs}
  \begin{split}
   &\frac{1}{\sqrt{2}}(|00\cdots0\rangle+|11\cdots1\rangle)\sqrt{p_{\text{even}}}|\text{even}\rangle_{\mu} \\
    &+\frac{1}{\sqrt{2}}(|00\cdots0\rangle-|11\cdots1\rangle)\sqrt{p_{\text{odd}}}|\text{odd}\rangle_{\mu},
  \end{split}
\end{equation}
where
\begin{equation}\label{coherentparity}
  \begin{split}
    |\text{even}\rangle_{\mu}& (|\text{odd}\rangle_{\mu})= \frac{1}{\sqrt{p_{\text{even}(\text{odd})}}}\sum_{n_1+n_2+\cdots+n_{N-1}\in \text{\textbf{even}}(\text{\textbf{odd}})}\\
    &\frac{\mathrm{e}^{-(N-1)\mu/2}(\mu)^{\frac{n_1+n_2+\cdots+n_{N-1}}{2}}}{n_1!n_2!\cdots n_{N-1}!}\{[\frac{p_1^{\dag}+(-1)^{i_1}p_{2L}^{\dag}}{\sqrt{2}}]^{n_1}\\
    &\otimes[\frac{(-1)^{i_1}p_{2R}^{\dag}+(-1)^{i_2}p_{3L}^{\dag}}{\sqrt{2}}]^{n_2}
       \otimes\cdots\\
    &\otimes[\frac{(-1)^{i_{N-2}}p_{(N-1)R}^{\dag}+(-1)^{i_{N-1}}p_{N}^{\dag}}{\sqrt{2}}]^{n_{N-1}}\}|\text{vac}\rangle,
  \end{split}
\end{equation}
\begin{equation}\label{peven}
\begin{split}
  p_{\text{even}}&=\sum_{n_1+n_2+\cdots+n_{N-1}\in \text{\textbf{even}}}\frac{\mathrm{e}^{-(N-1)\mu}(\mu)^{n_1+n_2+\cdots+n_{N-1}}}{n_1!n_2!\cdots n_{N-1}!},\\
  &=\mathrm{e}^{-(N-1)\mu}\cosh[(N-1)\mu]
\end{split}
\end{equation}
\begin{equation}\label{podd}
\begin{split}
  p_{\text{odd}}=1-p_{\text{even}}=\mathrm{e}^{-(N-1)\mu}\sinh[(N-1)\mu]
\end{split}
\end{equation}
with $i_1=i_2=\cdots=i_{N-1}=0$ for Eq.~\ref{Lclikcs}. Here, we omit the phase because Eve's announcement strategy cannot dependent on $\phi$.

When the phases are not announced, $|\text{even}\rangle_{\mu}(|\text{odd}\rangle_{\mu})$ is a mixture of photon number states from Eve's perspective after the phase randomization. While, if the phases are announced, $|\text{even}\rangle_{\mu}(|\text{odd}\rangle_{\mu})$ can no longer be regarded as a mixture of photon number states. Considering the Beam splitting attack using beam splitters with transmission rate $\eta$, the Eq.~\ref{Lclikcs} can be rewritten as
\begin{widetext}
  \begin{equation}\label{Lclikcs_BS_attack}
  \begin{split}
    &\frac{1}{\sqrt{2}}[|00\cdots 0\rangle|\sqrt{\frac{(1-\eta)\mu}{2}}\rangle|\sqrt{\frac{(1-\eta)\mu}{2}}\rangle\cdots|\sqrt{\frac{(1-\eta)\mu}{2}}\rangle\\
    &+|11\cdots 1\rangle|-\sqrt{\frac{(1-\eta)\mu}{2}}\rangle|-\sqrt{\frac{(1-\eta)\mu}{2}}\rangle\cdots|-\sqrt{\frac{(1-\eta)\mu}{2}}\rangle ]\sqrt{p_{\text{even}}^{\eta\mu}}|\text{even}\rangle_{\eta\mu}\\
    &+\frac{1}{\sqrt{2}}[|00\cdots0\rangle|\sqrt{\frac{(1-\eta)\mu}{2}}\rangle|\sqrt{\frac{(1-\eta)\mu}{2}}\rangle\cdots|\sqrt{\frac{(1-\eta)\mu}{2}}\rangle\\
    &-|11\cdots1\rangle|-\sqrt{(1-\eta)\frac{\mu}{2}}\rangle|-\sqrt{(1-\eta)\frac{\mu}{2}}\rangle\cdots|-\sqrt{(1-\eta)\frac{\mu}{2}}\rangle ]\sqrt{p_{\text{odd}}^{\eta\mu}}|\text{odd}\rangle_{\eta\mu}\\
    &=\frac{1}{\sqrt{2}}(|00\cdots0\rangle+|11\cdots1\rangle)[\sqrt{p_{\text{even}}^{(1-\eta)\mu}}|\text{even}\rangle_{(1-\eta)\mu}\sqrt{p_{\text{even}}^{\eta\mu}}|\text{even}\rangle_{\eta\mu}
    +\sqrt{p_{\text{odd}}^{(1-\eta)\mu}}|\text{odd}\rangle_{(1-\eta)\mu}\sqrt{p_{\text{odd}}^{\eta\mu}}|\text{odd}\rangle_{\eta\mu}]\\
    &+\frac{1}{\sqrt{2}}(|00\cdots0\rangle-|11\cdots1\rangle)[\sqrt{p_{\text{odd}}^{(1-\eta)\mu}}|\text{odd}\rangle_{(1-\eta)\mu}\sqrt{p_{\text{even}}^{\eta\mu}}|\text{even}\rangle_{\eta\mu}
    +\sqrt{p_{\text{even}}^{(1-\eta)\mu}}|\text{even}\rangle_{(1-\eta)\mu}\sqrt{p_{\text{odd}}^{\eta\mu}}|\text{odd}\rangle_{\eta\mu}],
  \end{split}
\end{equation}
\end{widetext}
where $|\text{even}\rangle_{\eta\mu}$ ($|\text{odd}\rangle_{\eta\mu}$) interferes before phase announcements, and cannot be used for eavesdropping. $\sqrt{p_{\text{odd}}^{\eta\mu}}$ is the normalization coefficient for pure state $|\text{odd}\rangle_{\eta\mu}$, and similar for other coefficients. While, $|\text{even}\rangle_{(1-\eta)\mu}$ and $|\text{odd}\rangle_{(1-\eta)\mu}$ are intercepted by Eve, and can not be decoupled from private qubits after phase announcements. Further, one can derive that
\begin{equation}\label{attack}
\begin{split}
   &\sqrt{p_{\text{even}}^{(1-\eta)\mu}}|\text{even}\rangle_{(1-\eta)\mu}\sqrt{p_{\text{even}}^{\eta\mu}}|\text{even}\rangle_{\eta\mu}\\
   &+\sqrt{p_{\text{odd}}^{(1-\eta)\mu}}|\text{odd}\rangle_{(1-\eta)\mu}\sqrt{p_{\text{odd}}^{\eta\mu}}|\text{odd}\rangle_{\eta\mu}\\
   &=\sqrt{p_{\text{even}}^{\mu}}|\text{even}\rangle_{\mu},\\
   &\sqrt{p_{\text{odd}}^{(1-\eta)\mu}}|\text{odd}\rangle_{(1-\eta)\mu}\sqrt{p_{\text{even}}^{\eta\mu}}|\text{even}\rangle_{\eta\mu}\\
   &+\sqrt{p_{\text{even}}^{(1-\eta)\mu}}|\text{even}\rangle_{(1-\eta)\mu}\sqrt{p_{\text{odd}}^{\eta\mu}}|\text{odd}\rangle_{\eta\mu}\\
   &= \sqrt{p_{\text{odd}}^{\mu}}|\text{odd}\rangle_{\mu}.
\end{split}
\end{equation}
From Eq.~\ref{phase_error}, the phase error is estimated for coherent states with intensity $\mu$. Meanwhile, as shown in Eq.~\ref{attack}, the phase error after the phase announcement can also be estimated by Eq.~\ref{phase_error}. This means that the phase error induced by the phase announcement has been estimated in Eq.~\ref{phase_error}, and can be corrected during the entanglement distillation protocol according to Eq.~\ref{distillation_supp}. Thus, the PM-QCC protocol is secure against the phase announcements.

Furthermore, supposing that we can estimate the phase error accurately in Eq.~\ref{phase_error}, the PM-QCC protocol is still secure even when the phase choices in the signal pulses are announced before Eve's measurement according to Eq.~\ref{attack}. Thus, the phase compensation method just provides a practical and secure way to align phases for signal pulses, and the PM-QCC protocol can be improved to a version without phase post-selection in signal pulses (see Appendix~\ref{without_PhaseSlice} for detail).

\section{Parameter Estimation for PM-QCC}\label{Parameter}
Experimentally, the overall gain $Q_{\mu}$ and marginal bit error rates (QBER) $E_{\mu,P_1P_2}^Z, E_{\mu,P_1P_3}^Z,\cdots, E_{\mu,P_1P_{N}}^Z$ can be directly estimated from the announced results. However, the phase error $E_{\mu}^X$ can not be measured directly in experiments. We can adopt the decoy-state method to estimate the phase error rate $E_{\mu}^X$.

As shown in Fig.~\ref{three_EDP_3}, supposing the phase matching condition $|j_m+j_m^a-j_{m+1}| \mod M=0$ or $M/2$ is satisfied, the click events in each branch are independent. In branch $1$, the gain $Q_{\mu,1}$ and the QBER $E_{\mu}^{Z,1}$ can be estimated as
\begin{equation}
\begin{split}
  Q_{\mu,1}&=P(D_{L_1})+P(D_{R_1}),\\
  E_{\mu}^{Z,1}&=\frac{P(D_{L_1})}{P(D_{L_1})+P(D_{R_1})},
\end{split}
\end{equation}
where $P(D_{L_1})$ means the probability that only detector $D_{L_1}$ clicks in branch $1$, $P(D_{R_1})$ means the probability that only the detector $D_{R_1}$ clicks in branch $1$. Because of the independence between the detection events of each branch when the phases are matched, the the gain and QBERs for other branches can be estimated as $Q_{\mu,t}=Q_{\mu,1}$ and $E_{\mu}^{Z,t}=E_{\mu}^{Z,1}$ ($2\leq t\leq N-1 $). Thus, the overall gain $Q_{\mu}$ and the marginal QBER $E_{\mu,P_1P_m}^Z$ between parties $P_1$ and $P_m$ are
\begin{subequations}
  \begin{equation}
  \begin{aligned}
    Q_{\mu}=(Q_{\mu,1})^{N-1}=p_{\text{odd}}Y_{\mu}^{\text{odd}}+p_{\text{even}}Y_{\mu}^{\text{even}},
  \end{aligned}\label{gain}
  \end{equation}
  \begin{equation}\label{marginal}
   E_{\mu,P_1P_m}^Z=\sum_{k=0}^{\lfloor \frac{m-2}{2}\rfloor} C_{m-1}^{2k+1}(E_{\mu}^{Z,1})^{2k+1}(1-E_{\mu}^{Z,1})^{m-2k-2},
   \end{equation}
\end{subequations}
where $C_{m-1}^{2k+1}$ is the Binomial coefficient, $\lfloor x \rfloor$ is the Floor function, $Y_{\mu}^{\text{odd}}$ ($Y_{\mu}^{\text{even}}$)is the overall yield for odd (even) photon number component. With the help of decoy-state method, we can estimate $p_{\text{odd}}Y_{\mu}^{\text{odd}}$ from Eq.~\ref{gain} (see Appendix~\ref{decoy} for detail). Then, the phase error rate $E_{\mu}^X$ can be estimated by Eq.~\ref{phase_N_party} and Eq.~\ref{phase_error}.

Without loss of generality, one can derive the gain $Q_{\mu,1}$ and the QBER $E_{\mu}^{Z,1}$ for branch 1~\cite{ma2018phase} when the phase matching condition $|j_1+j_1^a-j_{2}| \mod M=0$ or $M/2$ and $k_1=k_2=0$. Supposing the phase reference deviation $ \phi_0$ is constrained to $\phi_0 \in [-\frac{\pi}{M},\frac{\pi}{M})$ with the help of the phase-compensation method, the phase of parties $P_1$ and $P_2$ are uniformly distributed on $\phi_1\in[\frac{2\pi}{M}j_1,\frac{2\pi}{M}(j_1+1))$ and $\phi_2\in[\frac{2\pi}{M}j_1+\phi_0,\frac{2\pi}{M}(j_1+1)+\phi_0)$. To be simple and consistent with Ref.~\cite{ma2018phase}, we take $j_1=0$. Thus,
\begin{equation}\label{phase_compen}
  \begin{split}
    &\phi_1\in[0,\frac{2\pi}{M}),\\
    &\phi_2\in[\phi_0,\frac{2\pi}{M}+\phi_0).
  \end{split}
\end{equation}

As shown in  Fig.~\ref{evolution}, the evolution of the encoded state of parties $P_1$ and $P_2$ in branch 1 is
\begin{equation}\label{click}
  \begin{split}
    &|\mathrm{e}^{i\phi_1}\sqrt{\eta\mu/2}\rangle_{p_1}\otimes |\mathrm{e}^{i\phi_2}\sqrt{\eta\mu/2}\rangle_{p_{2L}}\\
    &\xrightarrow{\text{BS}}|\sqrt{\eta\mu/2}(\mathrm{e}^{i\phi_1}+\mathrm{e}^{i\phi_2})\rangle_{L_1}\otimes |\sqrt{\eta\mu/2}(\mathrm{e}^{i\phi_1}-\mathrm{e}^{i\phi_2})\rangle_{R_1}.
  \end{split}
\end{equation}
where the transmission efficiency $\eta$ consists of channel losses and detection efficiencies. From Eq.~\ref{click}, the pulses hitting detectors $D_{L_1}$ and $D_{R_1}$ are independent. The probabilities of click and non-click events for $D_{L_1}$ and $D_{R_1}$ can be directly calculated that
\begin{equation}\label{clickProbability}
  \begin{split}
    P(\bar{L}_1)&=(1-p_d)\exp(-\eta\mu\cos^2\frac{\phi_{\delta}}{2}),\\
     P(L_1)&=1- P(\bar{L}_1),\\
     P(\bar{R}_1)&=(1-p_d)\exp(-\eta\mu\sin^2\frac{\phi_{\delta}}{2}),\\
     P(R_1)&=1-P(\bar{R}_1).
  \end{split}
\end{equation}
where $P(L_1)$($P(\bar{L}_1)$) is the click (non-click) probability of detector $D_{L_1}$, $P(R_1)$($P(\bar{R}_1)$) is the click (non-click) probability of detector $D_{R_1}$ and $\phi_{\delta}=\phi_2-\phi_1$. The successful detection probabilities for branch 1 are,
\begin{equation}
  \begin{split}
    P(D_{L_1})&=P(L_1)P(\bar{R}_1),\\
     P(D_{R_1})&=P(\bar{L}_1)P(R_1),
  \end{split}
\end{equation}
respectively. Then, the gain $Q_{\mu,1}$ of branch1 is~\cite{ma2018phase}
\begin{equation}\label{gain_mu1}
\begin{split}
   Q_{\mu,1}&=P(D_{L_1})+P(D_{R_1})\\
            &\approx 1-\mathrm{e}^{-\eta\mu}+2p_d\mathrm{e}^{-\eta\mu},
\end{split}
\end{equation}
where the approximation is obtained by ignoring $\sin^2\frac{\phi_{\delta}}{2}$ with a small $\phi_{\delta}$ and ignoring the higher order term $p_d(1-\mathrm{e}^{-\eta\mu}+2p_d\mathrm{e}^{-\eta\mu})$. For given $\phi_{\delta}$, the QBER is~\cite{ma2018phase}
\begin{equation}
\begin{split}
  E_{\mu}^{Z,1}(\phi_{\delta})&=\frac{P(D_{R_1})}{P(D_{L_1})+P(D_{R_1})}\\
            &\approx \frac{\mathrm{e}^{-\eta\mu}}{Q_{\mu,1}} (p_d+\eta\mu\sin^2\frac{\phi_{\delta}}{2}).
\end{split}
\end{equation}
where the approximation is obtained by taking $\mathrm{e}^{\eta\mu\sin^2\frac{\phi_{\delta}}{2}}\approx 1+\eta\mu\sin^2\frac{\phi_{\delta}}{2}$ with a small $\phi_{\delta}$ and ignoring the higher order term $p_d(p_d+\eta\mu\sin^2\frac{\phi_{\delta}}{2})$. From Eq.~\ref{phase_compen} and $\phi_0 \in [-\frac{\pi}{M},\frac{\pi}{M})$, the QBER $E_{\mu}^{Z,1}$ is of the form
\begin{equation}\label{QBER}
\begin{split}
    E_{\mu}^{Z,1}&=\frac{M}{2\pi}\int_{-\frac{\pi}{M}}^{\frac{\pi}{M}}d\phi_0
    \int_{-\frac{3\pi}{M}}^{\frac{3\pi}{M}}d\phi_{\delta}f^{\phi_0}(\phi_{\delta})E_{\mu}^{Z,1}(\phi_{\delta})\\
    &=\frac{ (p_d+\eta\mu\mathrm{e}_{\delta})\mathrm{e}^{-\eta\mu}}{Q_{\mu,1}},
\end{split}
\end{equation}
where $\mathrm{e}_{\delta}=\frac{\pi}{M}-\frac{M^2}{\pi^2}\sin^3\frac{\pi}{M}$. Here, $f^{\phi_0}(\phi_{\delta})$ is the probability distribution of $\phi_{\delta}$ for given $\phi_0$,
\begin{equation*}
  f^{\phi_0}(\phi_{\delta})=\left\{
              \begin{array}{ll}
              (\frac{M}{2\pi})^2[\phi_{\delta}+(\frac{2\pi}{M}-\phi_0)],  \phi_{\delta}\in[\phi_0-\frac{2\pi}{M}, \phi_0), \\
              \\
              (\frac{M}{2\pi})^2[-\phi_{\delta}+(\frac{2\pi}{M}+\phi_0)], \phi_{\delta}\in[\phi_0, \phi_0+\frac{2\pi}{M}).
              \end{array}
            \right.
\end{equation*}
\section{Decoy states analysis for PM-QCC}\label{decoy}
As stated in the main text, the signal pulses with intensity $\mu$ are only used to estimate the gain $Q_{\mu}$ and marginal quantum bit error rates (QBER) $E_{\mu,P_1P_2}^Z$, $E_{\mu,P_1P_3}^Z$, $\cdots$, $E_{\mu,P_1P_{N}}^Z$. The phase error $E^X_{\mu}$ are estimated from decoy pulses in intensity set $\{\nu,\omega,\tau,\cdots,0\}$. The phase choices and intensities of the users are announced after Eve's announcement. Eve's attacks are independent of signal pulses and decoy pulses. Thus, the decoy states can by used to estimate the phase error $E^X_{\mu}$ in the signal pluses.

Considering the decoy pulses, the virtual sources in each branch of the protocol of Fig.~\ref{three_EDP_3} are simultaneously randomized if the phase-matching conditions are satisfied: $|\phi_{1}-\phi_{2}|=0~\text{or}~\pi$, $|\phi_2-\phi_3|=0~\text{or}~\pi$, $\cdots$, $|\phi_{N-1}-\phi_N|=0~\text{or}~\pi$. For simplicity, we take $N=3$ and $\phi_{1}=\phi_{2}=\cdots=\phi_{N}=\phi$, the virtual source under phase-matching condition is~\cite{zeng2019symmetry}
\begin{equation}
  \begin{split}
    \frac{1}{2\pi}\int_{0}^{2\pi}d\phi&|\mathrm{e}^{i\phi}\sqrt{\mu}\rangle_{C_1}|\mathrm{e}^{i\phi}\sqrt{\mu}\rangle_{C_2}\langle\mathrm{e}^{i\phi}\sqrt{\mu}|_{C_1}\langle\mathrm{e}^{i\phi}\sqrt{\mu}|_{C_2}\\
    &=\sum_{k}^{\infty}P_{2\mu}(k)| k \rangle\langle k|,
  \end{split}
\end{equation}
where $P_{2\mu}(k)=\mathrm{e}^{-2\mu}\frac{(2\mu)^{k}}{k!}$ is the probability of generating $k$ photons in the virtual source, $|k\rangle=\frac{[\frac{1}{\sqrt{2}}(C_1^{\dagger}+C_2^{\dagger})]^{k}}{\sqrt{k!}}|vac\rangle$ and $k=n_1+n_2$ is the total photon number of branch 1 and branch 2. Then the overall $Q_{\mu}$ (Eq.~\ref{gain}) and phase error rate $E_{\mu}^{X}$ (Eq.~\ref{phase_error}) are turn to be
\begin{subequations}
  \begin{equation}\label{gain_mu_gaussian}
  Q_{\mu}=\sum_{k}^{\infty}P_{2\mu}(k)\cdot Y_k,
  \end{equation}
\begin{equation}\label{Exupp}
  \begin{aligned}
    E_{\mu}^X&=\sum_{k\in\textbf{odd}}P_{2\mu}(k)\frac{Y_k}{Q_{\mu}},\\
             &=1-\sum_{k\in\textbf{even}}P_{2\mu}(k)\frac{Y_k}{Q_{\mu}},\\
             &=1-\mathrm{e}^{-2\mu}\cdot\frac{Y_0}{Q_{\mu}}-\mathrm{e}^{-2\mu}\frac{(2\mu)^2}{2}\cdot\frac{Y_2}{Q_{\mu}}-\cdots,\\
             &\leq 1-\mathrm{e}^{-2\mu}\cdot\frac{Y_0}{Q_{\mu}}-\mathrm{e}^{-2\mu}\frac{(2\mu)^2}{2}\cdot\frac{Y_2}{Q_{\mu}},\\
             &\leq E_{\mu}^{X,U}=1-\mathrm{e}^{-2\mu}\cdot\frac{Y_0}{Q_{\mu}}-\mathrm{e}^{-2\mu}\frac{(2\mu)^2}{2}\cdot\frac{Y_2^L}{Q_{\mu}},
  \end{aligned}
\end{equation}
\end{subequations}
where $Y_k \in [0,1]$ is the yield when $k$ photon are generated in the virtual source and $Y_2^L$ is the lower bound of the yield when 2 photons are generated in the virtual source. $E_{\mu}^{X,U}$ is the upper bound of the phase error rate. The first inequality is obtained by setting high-order terms including $Y_4$, $Y_6$, $\cdots$ to $0$. The second inequality is obtained by substituting $Y_2$ with its lower bound $Y_2^L$. From Eq.~\ref{gain_mu_gaussian}, one can obtain a set of overall gain with different decoy states which can be used to estimation the Yield $Y_k$. For 3-party PM-QCC protocol, four decoy states with intensities $\{\nu>\omega>\tau>0\}$  are adopted to estimate the $Y_2^L$ using the Gaussian elimination method~\cite{grasselli2019practical,Xu2013Practical}.
\begin{widetext}
\begin{equation}\label{gaussian}
  \begin{split}
    \mathrm{e}^{2\nu}Q_{\nu}&=Y_0+2\nu Y_1+\frac{(2\nu)^2}{2}Y_2+\frac{(2\nu)^3}{6}Y_3+\frac{(2\nu)^4}{4!}Y_4+\cdots,\\
    \mathrm{e}^{2\omega}Q_{\omega}&=Y_0+2\omega Y_1+\frac{(2\omega)^2}{2}Y_2+\frac{(2\omega)^3}{6}Y_3+\frac{(2\omega)^4}{4!}Y_4+\cdots,\\
    \mathrm{e}^{2\tau}Q_{\tau}&=Y_0+2\tau Y_1+\frac{(2\tau)^2}{2}Y_2+\frac{(2\tau)^3}{6}Y_3+\frac{(2\tau)^4}{4!}Y_4+\cdots,\\
    Q_{0}&=Y_0,
  \end{split}
\end{equation}
\end{widetext}
where $Q_{\nu}$, $Q_{\omega}$, $Q_{\tau}$ and $Q_{0}$ are overall gains for different decoy states. From Eq.~\ref{gaussian}, one can cancel out the terms of $Y_0$, $Y_1$ and $Y_3$ with the Gaussian elimination method and generate an equation given by
\begin{equation}\label{gaussian_1}
  G = G_2\cdot Y_2+G_4\cdot Y_4+ G_5\cdot Y_5+\cdots,
\end{equation}
where
\begin{widetext}
\begin{subequations}
  \begin{equation}
  \begin{split}
    G=&[2\omega(2\nu)^3-2\nu(2\omega)^3][2\tau(\mathrm{e}^{2\omega}Q_{\omega}-Q_0)-2\omega(\mathrm{e}^{2\tau}Q_{\tau}-Q_0)]\\
     &-[2\tau(2\omega)^3-2\omega(2\tau)^3][2\omega(\mathrm{e}^{2\nu}Q_{\nu}-Q_0)-2\nu(\mathrm{e}^{2\omega}Q_{\omega}-Q_0)],
  \end{split}
  \end{equation}
  \begin{equation}
    G_2=\frac{[2\omega(2\nu)^3-2\nu(2\omega)^3][2\tau(2\omega)^2-2\omega(2\tau)^2]-[2\tau(2\omega)^3-2\omega(2\tau)^3][2\omega(2\nu)^2-2\nu(2\omega)^2]}{2},
  \end{equation}
  \begin{equation}
    G_4=\frac{[2\omega(2\nu)^3-2\nu(2\omega)^3][2\tau(2\omega)^4-2\omega(2\tau)^4]-[2\tau(2\omega)^3-2\omega(2\tau)^3][2\omega(2\nu)^4-2\nu(2\omega)^4]}{4!},
  \end{equation}
  \begin{equation*}
    \vdots
  \end{equation*}
\end{subequations}
\end{widetext}
Since $\nu>\omega>\tau>0$, one can see that $G$, $G_2 >0$ while $G_4$, $G_5 \cdots < 0$ by simple calculation. Thus, the lower bound $Y_2^L$ is obtained by setting $Y_4=Y_5= \cdots =0$ from Eq.~\ref{gaussian_1} since $Y_k \in [0,1]$ (see~\cite{ma2008quantum} for reference),
\begin{widetext}
\begin{equation}\label{low_Y2} Y_2^L=\frac{2\{[2\omega(2\nu)^3-2\nu(2\omega)^3][2\tau(\mathrm{e}^{2\omega}Q_{\omega}-Q_0)-2\omega(\mathrm{e}^{2\tau}Q_{\tau}-Q_0)]-[2\tau(2\omega)^3-2\omega(2\tau)^3][2\omega(\mathrm{e}^{2\nu}Q_{\nu}-Q_0)-2\nu(\mathrm{e}^{2\omega}Q_{\omega}-Q_0)]\}}{[2\omega(2\nu)^3-2\nu(2\omega)^3][2\tau(2\omega)^2-2\omega(2\tau)^2]-[2\tau(2\omega)^3-2\omega(2\tau)^3][2\omega(2\nu)^2-2\nu(2\omega)^2]}.
\end{equation}
\end{widetext}
The upper bound of the phase error rate is
\begin{equation}
\begin{split}
 E_{\mu}^{X,U}=1-\mathrm{e}^{-2\mu}\cdot\frac{Y_0}{Q_{\mu}}-\mathrm{e}^{-2\mu}\frac{(2\mu)^2}{2}\cdot\frac{Y_2^L}{Q_{\mu}}.
\end{split}
\end{equation}
Thus, the lower bound of the key generation rate for 3-party PM-QCC is
\begin{equation}\label{keyrate3party}
  \begin{split}
    R_{3-\text{party}}&\geq R_{3-\text{party}}^L\\
    &= (\frac{2}{M})^{2} Q_{\mu}[1-f\cdot\max \{H(E_{\mu,P_1P_2}^Z),\\
    &H(E_{\mu,P_1P_3}^Z)\}-H(E_{\mu}^{X,U})].
  \end{split}
\end{equation}
%\begin{subequations}
%  \begin{equation}
%    a
%  \end{equation}
%  \begin{equation}
%    b
%  \end{equation}
%\end{subeuqations}

The above decoy-state method can be directly generalized to $N\geq 4$ parties PM-QCC protocols. When it comes to $N\geq 4$ parties, more linear constraints are needed to tightly estimate the phase error rate $E_{\mu}^X$ shown in Eq.~\ref{Exupp_N}. Specifically, based on the structure of the N-party GHZ state, there are more than two interference branches in the virtue protocol of Fig.~\ref{three_EDP_3}. If some of the branches have no photons, the corresponding yield $Y_k$ will be rather small. Thus, in order to obtain a tight upper bound of $E_{\mu}^X$ with Eq.~\ref{Exupp}, one need more decoy states to estimate the high-order terms of the yield $Y_k$ with the Gaussian elimination method shown from Eq.~\ref{gaussian} to \ref{low_Y2}. Here, we shown the main steps to conduct the presented decoy-state method as follows:

\begin{enumerate}
  \item[Step.1] For N-party PM-QCC protocol, the randomized virtual source under phase-matching condition $\phi_{1}=\phi_{2}=\cdots=\phi_{N}=\phi$ is
\begin{equation}
  \begin{split}
    \frac{1}{2\pi}\int_{0}^{2\pi}d\phi&|\mathrm{e}^{i\phi}\sqrt{\mu}\rangle_{C_1}\cdots|\mathrm{e}^{i\phi}\sqrt{\mu}\rangle_{C_{N-1}}\langle\mathrm{e}^{i\phi}\sqrt{\mu}|_{C_1}\cdots\langle\mathrm{e}^{i\phi}\sqrt{\mu}|_{C_{N-1}}\\
    &=\sum_{k}^{\infty}P_{(N-1)\mu}(k)| k \rangle\langle k|,
  \end{split}
\end{equation}
where $P_{(N-1)\mu}(k)=\mathrm{e}^{-(N-1)\mu}\frac{[(N-1)\mu]^{k}}{k!}$ is the probability of generating $k$ photons in the virtual source, $|k\rangle=\frac{[\frac{1}{\sqrt{N-1}}(C_1^{\dagger}+C_2^{\dagger}+\cdots+C_{N-1}^{\dagger})]^{k}}{\sqrt{k!}}|vac\rangle$ and $k=n_1+n_2+\cdots+n_{N-1}$ is the total photon number of all the branches.
  \item[Step.2] Then the overall $Q_{\mu}$ (Eq.~\ref{gain}) and phase error rate $E_{\mu}^{X}$ (Eq.~\ref{phase_error}) are turn to be
\begin{subequations}
  \begin{equation}\label{gain_mu_gaussian_N}
  Q_{\mu}=\sum_{k}^{\infty}P_{(N-1)\mu}(k)\cdot Y_k,
  \end{equation}
\begin{equation}\label{Exupp_N}
  \begin{aligned}
    E_{\mu}^X&=\sum_{k\in\textbf{odd}}P_{(N-1)\mu}(k)\frac{Y_k}{Q_{\mu}},\\
             &=1-\sum_{k\in\textbf{even}}P_{(N-1)\mu}(k)\frac{Y_k}{Q_{\mu}},\\
             &=1-\mathrm{e}^{-(N-1)\mu}\cdot\frac{Y_0}{Q_{\mu}}-\mathrm{e}^{-(N-1)\mu}\frac{[(N-1)\mu]^2}{2}\cdot\frac{Y_2}{Q_{\mu}}-\cdots,\\
             &\leq 1-\mathrm{e}^{-(N-1)\mu}\cdot\frac{Y_0}{Q_{\mu}}-\mathrm{e}^{-(N-1)\mu}\frac{[(N-1)\mu]^2}{2}\cdot\frac{Y_2}{Q_{\mu}}-\cdots\\
              &-\mathrm{e}^{-(N-1)\mu}\frac{[(N-1)\mu]^{N_{\text{cut}}}}{N_{\text{cut}}!}\cdot \frac{Y_{N_{\text{cut}}}}{Q_{\mu}},\\
             &\leq E_{\mu}^{X,U}=1-\mathrm{e}^{-(N-1)\mu}\cdot\frac{Y_0}{Q_{\mu}}-\mathrm{e}^{-(N-1)\mu}\frac{[(N-1)\mu]^2}{2}\cdot\frac{Y_2^L}{Q_{\mu}}-\cdots\\
             &-\mathrm{e}^{-(N-1)\mu}\frac{[(N-1)\mu]^{N_{\text{cut}}}}{N_{\text{cut}}!}\cdot \frac{Y^L_{N_{\text{cut}}}}{Q_{\mu}},
  \end{aligned}
\end{equation}
\end{subequations}
where $N_{\text{cut}}$ is the cut number to bound the phase error rate, and $N_{\text{cut}}=N-1$ ($N$) if $N$ is odd (even)

\item[Step.3] The Gaussian elimination method shown in Eq.~\ref{gaussian} to \ref{low_Y2} is adopt to estimate the lower bounds $Y_2^L, \cdots, Y_{N_{\text{cut}}}^L$. To estimate the high-order terms $Y_{k}^L$, one only need to add extra linear constraints to Eq.~\ref{gaussian} by adding extra decoy states, say $\{\nu,\omega,\tau,\cdots,0\}$, to construct Eq.~\ref{gaussian_1}. Then, one can directly obtain the high-order terms $Y_{k}^L$ from Eq.~\ref{gaussian_1}.

\item[Step.4] With the lower bounds $Y_2^L, \cdots, Y_{N_{\text{cut}}}^L$, one can obtain the upper bound of the phase error rate $E_{\mu}^{X,U}$ according to Eq.~\ref{Exupp_N}.
  \end{enumerate}

Definitely, from the view of linear program, to well bound the yield $Y_{N_{\text{cut}}}$, $N_{\text{cut}}+1$ linear constraints is needed using the Gaussian elimination method. Thus, the number of decoy sates increases linearly with the communication parties $N$ in the presented decoy-state method. On the other hand, if one adopts the decoy-state estimation using the data when different parties send out different intensities, one can generate more decoy-state constraints of Eq.~\ref{gaussian}. In this case, when the number of parties N gets larger, the number of constraints also increases. This has already been studied in quantum key distribution, for example, see~\cite{curty2019simple,grasselli2019practical}. From this point of view, when N gets larger, one may not need more decoy states for each party. We remark that, similar problems have been solved in a different scenario in~\cite{yuan2016simulating} where few decoy-state settings are enough to generate a good estimation of the yield that each of the N parties send out one photon. Thus, with more advanced decoy-state method, it is possible to reduce the required decoy-state number for each party.

\section{PM-QCC* without phase post-selection in the signal pulses}\label{without_PhaseSlice}
As stated in Appendix~\ref{PMQCC}, supposing that we can estimate the phase error accurately, the PM-QCC protocol is still secure even when the phase choices in the signal pulses are announced before Eve's measurement. Without loss of generality, the phase choices for signal pulses of party $P_i$ can be set to $\phi_{i}=0$. The PM-QCC* without phase post-selection in the signal pulses run as follows:
\begin{enumerate}
  \item[Step.1] \textbf{Preparation}: The pulses are divided to be signal mode with intensity $\mu$ and decoy mode with intensities $\{\nu,\omega,\tau,\cdots,0\}$. In the signal mode, Party $P_1$ randomly generates one bit $k_{1}\in\{0,1\}$ and a coherent pulse $|\sqrt{\mu_1}\rangle$. Then, he encodes the random bit to the coherent pulse and get a encoded coherent pulses $|\mathrm{e}^{i\pi k_{1}}\sqrt{\mu_1}\rangle$. Similarly, parties $P_2$, $\cdots$, $P_N$ get encoded coherent pulses $|\mathrm{e}^{i\pi k_{2}}\sqrt{\mu_{2}}\rangle$, $\cdots$, $|\mathrm{e}^{i\pi k_{N}}\sqrt{\mu_{N}}\rangle$, respectively. In the decoy mode, party $P_1$ randomly generates one bit $k_{1}\in\{0,1\}$ and a coherent pulse with random phase $\phi_{1}\in[0,2\pi)$. Then, he encodes the random bit to the coherent pulse and get a phase randomized coherent pulses $|\mathrm{e}^{i(\phi_{1}+\pi k_{1})}\sqrt{\mu_1}\rangle$. Similarly, parties $P_2$, $\cdots$, $P_N$ prepare their phase randomized coherent pulses $|\mathrm{e}^{i(\phi_{2}+\pi k_{2})}\sqrt{\mu_{2}}\rangle$, $\cdots$, $|\mathrm{e}^{i(\phi_{N}+\pi k_{N})}\sqrt{\mu_{N}}\rangle$, respectively.

      As shown in Fig.~\ref{PMQCC_setup}, the experimental setup is asymmetric for parties $P_1$, $P_N$ and $P_2$, $\cdots$, $P_{N-1}$. Thus, the intensities of the weak coherent pulses for parties $P_1$, $P_N$ are set to be $\mu_1,\mu_N \in\{\frac{\mu}{2}>\frac{\nu}{2}>\frac{\omega}{2}>\frac{\tau}{2}>\cdots>0\}$, while the intensities for parties $P_2$, $\cdots$, $P_{N-1}$ are set to be $\mu_t \in\{\mu>\nu>\omega>\tau>\cdots>0\}$ ($2\leq t\leq N-1$) for signal and decoy pulses. The pulses corresponding with intensity $\mu$ are signal pulses and the pulses corresponding with intensities $\{\nu,\omega,\tau,\cdots,0\}$ are used as decoy pulses.

  \item[Step.2] \textbf{Measurement}: Same as PM-QCC.
  \item[Step.3] \textbf{Announcement}: Same as PM-QCC.

  \item[Step.4] \textbf{Sifting}: Same as PM-QCC.
  \item[Step.5] \textbf{Parameter estimation and Key distillation}: The above steps are repeated enough times to distill the raw key bits. From the data set generated by the signal pulses, the users can directly estimate the gain $Q_{\mu}^*$ and marginal quantum bit error rates (QBER) $E_{\mu,P_1P_2}^{Z*}$, $E_{\mu,P_1P_3}^{Z*}$, $\cdots$, $E_{\mu,P_1P_{N}}^{Z*}$ from the measurement results. From the data set generated by the decoy pulses, the users can estimate the phase error $E^{X*}_{\mu}$ according to decoy-state methods. Finally, they distill private key bits by performing error correction and privacy amplification on the raw key.

  \end{enumerate}

According to Appendix~\ref{decoy}, the decoy states are simultaneously randomized under phase-matching condition, and the above mentioned decoy-state method can be directly used to estimate the phase error $E_{\mu}^{X}$ in signal mode for the PM-QCC protocol. However, in the PM-QCC* protocol the signal pulses from the parties can no longer be regarded as photon number states since the phase randomization for signal pulses has been cancelled out. Thus, the above decoy states discussion becomes unsuitable for the PM-QCC* protocol, and more delicate decoy-state method is required to evaluate the phase error rate in the signals~\cite{lin2018simple,curty2019simple,maeda2019repeaterless}. For example, as in~\cite{curty2019simple}, the estimation of phase error rate is converted to the estimation of the yields for the photon number state, which can be estimated using phase randomized decoy states. Thus, the phase randomized decoy states with different intensities can in principle be used to constraint the phase error rate $E_{\mu}^X$ tightly and we leave if for further studies. The gain $Q_{\mu,1}^*$ of branch1 is
\begin{equation}\label{gain_mu1*}
\begin{split}
   Q_{\mu,1}^*=Q_{\mu,1},
\end{split}
\end{equation}
the QBER is
\begin{equation}
\begin{split}
  E_{\mu}^{Z*,1}&=\frac{P(D_{R_1})}{P(D_{L_1})+P(D_{R_1})}\\
            &=\frac{(1 - p_d)(\mathrm{e}^{-\eta\mu(1-e_{\delta})})\cdot(1-(1-pd)(\mathrm{e}^{-\eta\mu e_{\delta}}))}{Q_{\mu,1}^*},
\end{split}
\end{equation}
where $e_{\delta}$ is the phase misaligned error for signal mode during the phase reference.
From the Eq.~\ref{gain}, Eq.~\ref{marginal} and Eq.~\ref{phase_error}, the overall gain, Marginal QBERs and phase error are
\begin{subequations}
  \begin{equation}
  \begin{aligned}
    Q_{\mu}^*=Q_{\mu},
  \end{aligned}\label{}
  \end{equation}
  \begin{equation}\label{}
   E_{\mu,P_1P_m}^{Z*}=\sum_{k=0}^{\lfloor \frac{m-2}{2}\rfloor} C_{m-1}^{2k+1}(E_{\mu}^{Z*,1})^{2k+1}(1-E_{\mu}^{Z*,1})^{m-2k-2},
   \end{equation}
   \begin{equation}
  \begin{aligned}
    E_{\mu}^{X*}=E_{\mu}^{X}.
  \end{aligned}\label{}
  \end{equation}
\end{subequations}
According to a former discussion on PM-QKD~\cite{lin2018simple}, one can simply suppose that a sufficient parameter estimation can be made by a complete characterization of Eve's measurement operators. As a result, the estimated phase error rate $E_{\mu}^{X*}$ is equal to the detected fraction of the odd photon number component in the infinite-key regime. Then, the key generation rate for PM-QCC protocol without phase-matching condition in signal modes is
\begin{equation}\label{keyratewithouphasematching}
  \begin{split}
    R_{N-\text{party}}^*=&Q_{\mu}^*[1-f\cdot\max \{H(E_{\mu,P_1P_2}^{Z*}), \\
                       &H(E_{\mu,P_1P_3}^{Z*}), \cdots, H(E_{\mu,P_1P_{N}}^{Z*})\}-H(E_{\mu}^{X*})].
  \end{split}
  \end{equation}

\section{Comparison with MDI-QCC}\label{comparison}
We compare the performance of PM-QCC with that of MDI-QCC in Ref.~\cite{fu2015long} when $N=3$. The key rate of MDI-QCC in simulation is
\begin{equation}\label{MDIKeyRate}
  R_{MDI-QCC} = Q_{111}^Z[1-H(e_{111}^{BX})]-H(E_{\mu\nu\omega}^{Z*}) f Q_{\mu\nu\omega}^Z,
\end{equation}
where $E_{\mu\nu\omega}^{Z*}=\max\{ H(E_{\mu\nu\omega}^{ZAB}), H(E_{\mu\nu\omega}^{ZAC}) \}$, $Q_{\mu\nu\omega}^Z(E_{\mu\nu\omega}^{Z*})$ is the gain (QBER) of the Z basis, $Q_{111}^Z$ is the gain of single photon component, $e_{111}^{BX}$ is the single photon QBER of the X basis, $f$ is the error correction efficiency and $H(x)=-x\log_2(x)-(1-x)\log_2(1-x)$ is the binary entropy function.

From Eq.~\ref{gain}, one can obtain that $Q_{\mu}\varpropto \eta^{N-1}$, where $\eta=\mathrm{e}^{-\alpha L/10}$ is the transmission rate of the optical channel, $\alpha$ is the corresponding loss rate, $L$ is the distance from each party to the measurement station. Thus, the key generate rate $R_{PM-QCC} \varpropto \eta^{N-1}$. Nevertheless, in measurement device independent quantum cryptographic conferencing (MDI-QCC)~\cite{fu2015long}, one can calculate that $ Q_{111}^Z \varpropto \eta^{3}$ when dark-count rate $p_d=0$. Intuitively, $N$ photons are coincidentally detected by $N$ different detectors for $N$-party MDI-QCC networks based on the GHZ state analyzer~\cite{qian2005universal} with coincidence probability $P_{\text{co}}\varpropto \eta^N$. Then, we obtain that $R_{MDI-QCC} \varpropto \eta^{N} $ according to Eq.~\ref{MDIKeyRate}. Therefore, the presented PM-QCC can improve the key generation rate from $\mathrm{O}(\eta^{N})$ to $\mathrm{O}(\eta^{N-1})$.

\begin{figure}[htbp]
  \centering
   \subfigure[]{
   \label{reduce_small}
  \includegraphics[width=0.3\textwidth]{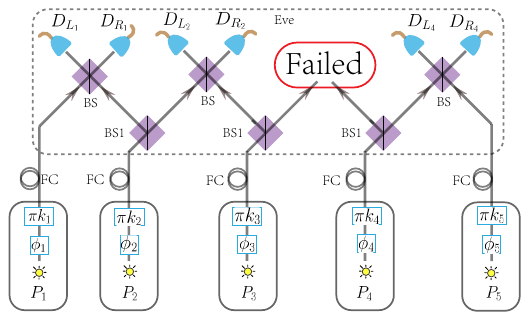}
  }
  \subfigure[]{
  \label{small_scale_PM_QCC_EDP}
  \includegraphics[width=0.23\textwidth]{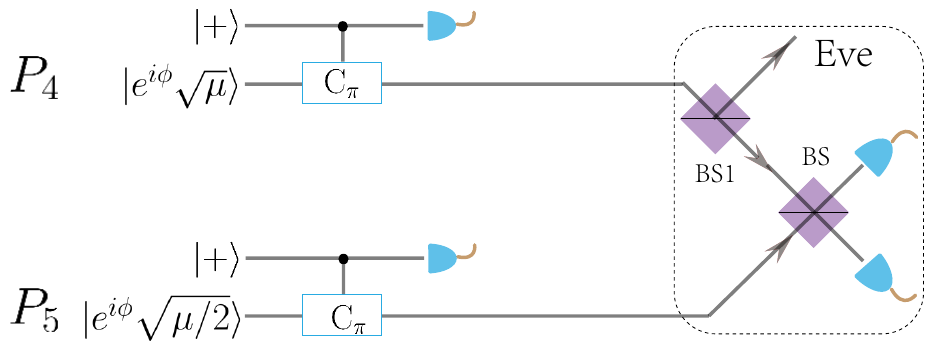}
  }
  \subfigure[]{
  \label{small_scale_three_EDP_3}
  \includegraphics[width=0.23\textwidth]{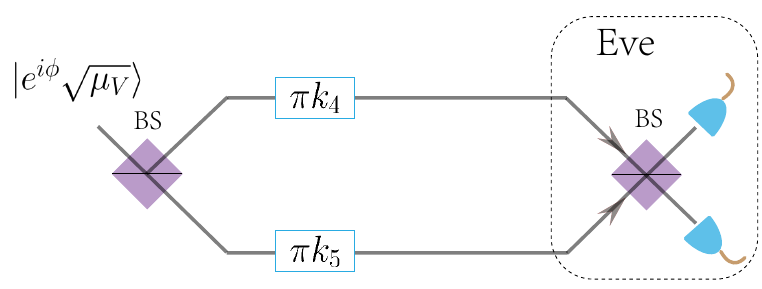}
  }
  \caption{(a) The 3-party PM-QCC network and the 2-party PM-QKD protocol reduced from the 5-party PM-QCC network. In some rounds of 5-party PM-QCC only sets $\{P_1, P_2, P_5\}$ and $\{P_4, P_5\}$ are successfully interfered, while, the interference of $\{P_3, P_4\}$ failed. (b) The entanglement based PM-QKD protocol for parties $P_4$ and $P_5$. (c) The equivalent entanglement based PM-QKD protocol for $P_4$ and $P_5$ with virtual source. BS: Beam Splitter. FC: Fiber Channel. $\text{C}_{\pi}$: the control phase gate.}
\end{figure}
\section{Reduce to Small-scale PM-QCC Networks}\label{small}
Practically, the $N$-party PM-QCC network might not work perfectly, i.e. there might not be perfect $N$-party interferences of the weak coherent states. We consider the case that $N'$-party inferences ($N'$ parties are near-neighbor connected, and $2\leq N' \leq N$) are realized in some rounds of $N$-party PM-QCC network. For example,  as shown in Fig.~\ref{reduce_small}, only the interferences of sets $\{P_1, P_2, P_3\}$ and $\{P_4,P_5\}$ are realized, while, the interference of $\{P_3, P_4\}$ failed in the $5$-party PM-QCC network. According to Step.1 of the PM-QCC network, the coherent pulses sent by parties at the broken point are $|\sqrt{\mu}\rangle$ instead of $|\sqrt{\mu/2}\rangle$ (here,we omit the phase information). Considering asymmetric mean photon numbers for two arms of an interference branch, we show that reduced small-scale PM-QCC networks and PM-QKD protocol can also be realized securely.

\begin{table}[h]
  \caption{The performance for the reduced $3$-party PM-QCC network shown in Fig.~\ref{reduce_small}. The key generation rate $R$, mean photon number $\mu$ and phase slice number $M$ are optimized with $p_d=7.2\times10^{-8}$, $\eta_d=65\%$, $f=1.16$ and $\alpha=0.2~\mathrm{dB}/\mathrm{km}$ at different transmission distance $L$.}\label{table:reduced_simulation}
  \begin{tabular*}{0.45\textwidth}{@{\extracolsep{\fill}} c c c c  }
  \hline\hline
  $R_{\text{reduced 3-party}}~(bits~per~pulse)$ &$L~(\text{Km})$ & $\mu$ & $M$ \\
  \hline
  $1.7060\times10^{-7}$ & $50$ & $0.1059$ & $13$ \\
  $1.6152\times10^{-9}$ & $100$ & $0.1032$ & $13$\\
  \hline\hline
 \end{tabular*}
\end{table}

Without loss of generality, we consider the interference of parties $P_4$ and $P_5$ as shown in Fig.~\ref{small_scale_PM_QCC_EDP}. The virtual entanglement based PM-QKD protocol with a virtual weak coherent state source $|\mathrm{e}^{i\phi}\sqrt{\mu_\text{V}}\rangle$ ($\mu_\text{V}=\mu_\text{a}+\mu_\text{b}$) is shown in Fig.~\ref{small_scale_three_EDP_3}. For the above small-scale PM-QCC network and PM-QKD protocol, the average number of photons for parties at the boundary are not equal. The weak coherent pulses after encoding of parties $P_4$ and $P_5$ are $|\mathrm{e}^{i\phi_a}\sqrt{\mu_a}\rangle$ and $|\mathrm{e}^{i\phi_b}\sqrt{\mu_b}\rangle$, respectively, with $\mu_a=\mu$ and $\mu_b=\frac{\mu}{2}$. As shown in Fig.~\ref{small_scale_PM_QCC_EDP}, the weak coherent states arriving for interference are equal, i.e. $|\mathrm{e}^{i\phi_a}\sqrt{\eta_a\mu_a}\rangle$ and $|\mathrm{e}^{i\phi_b}\sqrt{\eta_b\mu_b}\rangle$ with $\eta_a\mu_a=\eta_b\mu_b=\frac{\eta\mu}{2}$. According to Eq.~\ref{click}~$\sim$~\ref{QBER}, the gain and QBER of this  branch with asymmetric average number of photons are equal to the original branch.
\begin{equation}\label{reduce_one_branch}
  \begin{split}
    Q_{\mu_{\text{V}},1}^{'} &= Q_{\mu,1}\\
    E_{\mu_{\text{V}},1}^{'Z,1} &= E_{\mu}^{Z,1}.
  \end{split}
\end{equation}
Meanwhile, higher amounts of weak coherent pulses are lost during the transmission. This will result in a lower key generation rate for small-scale PM-QCC network or PM-QKD protocol, which is embodied in the estimation of $E_{\mu}^X$ from Eq.~\ref{phase_error}. Specifically, we present the performance of the reduced $3$-party PM-QCC network from $5$-party PM-QCC network shown in Fig.~\ref{reduce_small} in Table.~\ref{table:reduced_simulation}. In practice, the coherent pulses and phase slices $M$ might be pre-optimized for bigger PM-QCC network, and the key generation rate for the reduced $3$-party PM-QCC network might decrease to a lower level compared with the results in Table.~\ref{table:reduced_simulation}. However, according to Eq.~\ref{reduce_one_branch}, the gain of this asymmetric branch still scales with $Q_{\mu_{\text{V}},1}^{'}\varpropto \eta$. Thus, a key generation rate scales with $R_{\text{reduced}~PM-QCC} \varpropto \eta^{N'-1}$ can also be obtained according to Eq.~\ref{distillation_supp} for the $N'$ parties PM-QCC network.

Therefore, the $N$-party PM-QCC network can be reduced to small-scale PM-QCC networks and PM-QKD protocols when there are only parts of the $N$ parties are interfered altogether.

\end{appendix}

%%%%%%%%%%%%%%%%%%%%%%%%%%%%%%%%%%%%%%%%
% choose a style
%\bibliographystyle{ieeetr}
%\bibliographystyle{unsrt}
%%%%%%%%%%%%%%%%%%%%%%%%%%%%%%%%%%%%%%%%

%%%%%%%%%%%%%%%%%%%%%%%%%%%%%%%%%%%%%%%%
 %choose a .bib file
%\bibliographystyle{apsrev4-1}
%\bibliography{PMQCC}

\begin{thebibliography}{61}%
\makeatletter
\providecommand \@ifxundefined [1]{%
 \@ifx{#1\undefined}
}%
\providecommand \@ifnum [1]{%
 \ifnum #1\expandafter \@firstoftwo
 \else \expandafter \@secondoftwo
 \fi
}%
\providecommand \@ifx [1]{%
 \ifx #1\expandafter \@firstoftwo
 \else \expandafter \@secondoftwo
 \fi
}%
\providecommand \natexlab [1]{#1}%
\providecommand \enquote  [1]{``#1''}%
\providecommand \bibnamefont  [1]{#1}%
\providecommand \bibfnamefont [1]{#1}%
\providecommand \citenamefont [1]{#1}%
\providecommand \href@noop [0]{\@secondoftwo}%
\providecommand \href [0]{\begingroup \@sanitize@url \@href}%
\providecommand \@href[1]{\@@startlink{#1}\@@href}%
\providecommand \@@href[1]{\endgroup#1\@@endlink}%
\providecommand \@sanitize@url [0]{\catcode `\\12\catcode `\$12\catcode
  `\&12\catcode `\#12\catcode `\^12\catcode `\_12\catcode `\%12\relax}%
\providecommand \@@startlink[1]{}%
\providecommand \@@endlink[0]{}%
\providecommand \url  [0]{\begingroup\@sanitize@url \@url }%
\providecommand \@url [1]{\endgroup\@href {#1}{\urlprefix }}%
\providecommand \urlprefix  [0]{URL }%
\providecommand \Eprint [0]{\href }%
\providecommand \doibase [0]{http://dx.doi.org/}%
\providecommand \selectlanguage [0]{\@gobble}%
\providecommand \bibinfo  [0]{\@secondoftwo}%
\providecommand \bibfield  [0]{\@secondoftwo}%
\providecommand \translation [1]{[#1]}%
\providecommand \BibitemOpen [0]{}%
\providecommand \bibitemStop [0]{}%
\providecommand \bibitemNoStop [0]{.\EOS\space}%
\providecommand \EOS [0]{\spacefactor3000\relax}%
\providecommand \BibitemShut  [1]{\csname bibitem#1\endcsname}%
\let\auto@bib@innerbib\@empty
%</preamble>
\bibitem [{\citenamefont {Elliott}(2002)}]{Elliott_2002Building}%
  \BibitemOpen
  \bibfield  {author} {\bibinfo {author} {\bibfnamefont {C.}~\bibnamefont
  {Elliott}},\ }\href@noop {} {\bibfield  {journal} {\bibinfo  {journal} {New
  J. Phys.}\ }\textbf {\bibinfo {volume} {4}},\ \bibinfo {pages} {46} (\bibinfo
  {year} {2002})}\BibitemShut {NoStop}%
\bibitem [{\citenamefont {Elliott}\ \emph {et~al.}(2005)\citenamefont
  {Elliott}, \citenamefont {Colvin}, \citenamefont {Pearson}, \citenamefont
  {Pikalo}, \citenamefont {Schlafer},\ and\ \citenamefont
  {Yeh}}]{elliott2005current}%
  \BibitemOpen
  \bibfield  {author} {\bibinfo {author} {\bibfnamefont {C.}~\bibnamefont
  {Elliott}}, \bibinfo {author} {\bibfnamefont {A.}~\bibnamefont {Colvin}},
  \bibinfo {author} {\bibfnamefont {D.}~\bibnamefont {Pearson}}, \bibinfo
  {author} {\bibfnamefont {O.}~\bibnamefont {Pikalo}}, \bibinfo {author}
  {\bibfnamefont {J.}~\bibnamefont {Schlafer}}, \ and\ \bibinfo {author}
  {\bibfnamefont {H.}~\bibnamefont {Yeh}},\ }\href@noop {} {\bibfield
  {journal} {\bibinfo  {journal} {arXiv: quant-ph/0503058}\ } (\bibinfo {year}
  {2005})}\BibitemShut {NoStop}%
\bibitem [{\citenamefont {Peev}\ \emph {et~al.}(2009)\citenamefont {Peev},
  \citenamefont {Pacher}, \citenamefont {All{\'{e}}aume}, \citenamefont
  {Barreiro}, \citenamefont {Bouda}, \citenamefont {Boxleitner}, \citenamefont
  {Debuisschert}, \citenamefont {Diamanti}, \citenamefont {Dianati},
  \citenamefont {Dynes} \emph {et~al.}}]{Peev_2009The}%
  \BibitemOpen
  \bibfield  {author} {\bibinfo {author} {\bibfnamefont {M.}~\bibnamefont
  {Peev}}, \bibinfo {author} {\bibfnamefont {C.}~\bibnamefont {Pacher}},
  \bibinfo {author} {\bibfnamefont {R.}~\bibnamefont {All{\'{e}}aume}},
  \bibinfo {author} {\bibfnamefont {C.}~\bibnamefont {Barreiro}}, \bibinfo
  {author} {\bibfnamefont {J.}~\bibnamefont {Bouda}}, \bibinfo {author}
  {\bibfnamefont {W.}~\bibnamefont {Boxleitner}}, \bibinfo {author}
  {\bibfnamefont {T.}~\bibnamefont {Debuisschert}}, \bibinfo {author}
  {\bibfnamefont {E.}~\bibnamefont {Diamanti}}, \bibinfo {author}
  {\bibfnamefont {M.}~\bibnamefont {Dianati}}, \bibinfo {author} {\bibfnamefont
  {J.~F.}\ \bibnamefont {Dynes}},  \emph {et~al.},\ }\href@noop {} {\bibfield
  {journal} {\bibinfo  {journal} {New J. Phys.}\ }\textbf {\bibinfo {volume}
  {11}},\ \bibinfo {pages} {075001} (\bibinfo {year} {2009})}\BibitemShut
  {NoStop}%
\bibitem [{\citenamefont {Xu}\ \emph {et~al.}(2009)\citenamefont {Xu},
  \citenamefont {Chen}, \citenamefont {Wang}, \citenamefont {Yin},
  \citenamefont {Zhang}, \citenamefont {Liu}, \citenamefont {Zhou},
  \citenamefont {Zhao}, \citenamefont {Li}, \citenamefont {Liu} \emph
  {et~al.}}]{Xu2009Field}%
  \BibitemOpen
  \bibfield  {author} {\bibinfo {author} {\bibfnamefont {F.}~\bibnamefont
  {Xu}}, \bibinfo {author} {\bibfnamefont {W.}~\bibnamefont {Chen}}, \bibinfo
  {author} {\bibfnamefont {S.}~\bibnamefont {Wang}}, \bibinfo {author}
  {\bibfnamefont {Z.}~\bibnamefont {Yin}}, \bibinfo {author} {\bibfnamefont
  {Y.}~\bibnamefont {Zhang}}, \bibinfo {author} {\bibfnamefont
  {Y.}~\bibnamefont {Liu}}, \bibinfo {author} {\bibfnamefont {Z.}~\bibnamefont
  {Zhou}}, \bibinfo {author} {\bibfnamefont {Y.}~\bibnamefont {Zhao}}, \bibinfo
  {author} {\bibfnamefont {H.}~\bibnamefont {Li}}, \bibinfo {author}
  {\bibfnamefont {D.}~\bibnamefont {Liu}},  \emph {et~al.},\ }\href@noop {}
  {\bibfield  {journal} {\bibinfo  {journal} {Chin. Sci. Bul.}\ }\textbf
  {\bibinfo {volume} {54}},\ \bibinfo {pages} {2991} (\bibinfo {year}
  {2009})}\BibitemShut {NoStop}%
\bibitem [{\citenamefont {Stucki}\ \emph {et~al.}(2011)\citenamefont {Stucki},
  \citenamefont {Legr{\'{e}}}, \citenamefont {Buntschu}, \citenamefont
  {Clausen}, \citenamefont {Felber}, \citenamefont {Gisin}, \citenamefont
  {Henzen}, \citenamefont {Junod}, \citenamefont {Litzistorf}, \citenamefont
  {Monbaron} \emph {et~al.}}]{Stucki_2011Long}%
  \BibitemOpen
  \bibfield  {author} {\bibinfo {author} {\bibfnamefont {D.}~\bibnamefont
  {Stucki}}, \bibinfo {author} {\bibfnamefont {M.}~\bibnamefont {Legr{\'{e}}}},
  \bibinfo {author} {\bibfnamefont {F.}~\bibnamefont {Buntschu}}, \bibinfo
  {author} {\bibfnamefont {B.}~\bibnamefont {Clausen}}, \bibinfo {author}
  {\bibfnamefont {N.}~\bibnamefont {Felber}}, \bibinfo {author} {\bibfnamefont
  {N.}~\bibnamefont {Gisin}}, \bibinfo {author} {\bibfnamefont
  {L.}~\bibnamefont {Henzen}}, \bibinfo {author} {\bibfnamefont
  {P.}~\bibnamefont {Junod}}, \bibinfo {author} {\bibfnamefont
  {G.}~\bibnamefont {Litzistorf}}, \bibinfo {author} {\bibfnamefont
  {P.}~\bibnamefont {Monbaron}},  \emph {et~al.},\ }\href@noop {} {\bibfield
  {journal} {\bibinfo  {journal} {New J. Phys.}\ }\textbf {\bibinfo {volume}
  {13}},\ \bibinfo {pages} {123001} (\bibinfo {year} {2011})}\BibitemShut
  {NoStop}%
\bibitem [{\citenamefont {Kimble}(2008)}]{kimble2008quantum}%
  \BibitemOpen
  \bibfield  {author} {\bibinfo {author} {\bibfnamefont {H.~J.}\ \bibnamefont
  {Kimble}},\ }\href@noop {} {\bibfield  {journal} {\bibinfo  {journal}
  {Nature}\ }\textbf {\bibinfo {volume} {453}},\ \bibinfo {pages} {1023}
  (\bibinfo {year} {2008})}\BibitemShut {NoStop}%
\bibitem [{\citenamefont {Liao}\ \emph {et~al.}(2018)\citenamefont {Liao},
  \citenamefont {Cai}, \citenamefont {Handsteiner}, \citenamefont {Liu},
  \citenamefont {Yin}, \citenamefont {Zhang}, \citenamefont {Rauch},
  \citenamefont {Fink}, \citenamefont {Ren}, \citenamefont {Liu} \emph
  {et~al.}}]{Liao2018Satellite}%
  \BibitemOpen
  \bibfield  {author} {\bibinfo {author} {\bibfnamefont {S.-K.}\ \bibnamefont
  {Liao}}, \bibinfo {author} {\bibfnamefont {W.-Q.}\ \bibnamefont {Cai}},
  \bibinfo {author} {\bibfnamefont {J.}~\bibnamefont {Handsteiner}}, \bibinfo
  {author} {\bibfnamefont {B.}~\bibnamefont {Liu}}, \bibinfo {author}
  {\bibfnamefont {J.}~\bibnamefont {Yin}}, \bibinfo {author} {\bibfnamefont
  {L.}~\bibnamefont {Zhang}}, \bibinfo {author} {\bibfnamefont
  {D.}~\bibnamefont {Rauch}}, \bibinfo {author} {\bibfnamefont
  {M.}~\bibnamefont {Fink}}, \bibinfo {author} {\bibfnamefont {J.-G.}\
  \bibnamefont {Ren}}, \bibinfo {author} {\bibfnamefont {W.-Y.}\ \bibnamefont
  {Liu}},  \emph {et~al.},\ }\href@noop {} {\bibfield  {journal} {\bibinfo
  {journal} {Phys. Rev. Lett.}\ }\textbf {\bibinfo {volume} {120}},\ \bibinfo
  {pages} {030501} (\bibinfo {year} {2018})}\BibitemShut {NoStop}%
\bibitem [{\citenamefont {Wehner}\ \emph {et~al.}(2018)\citenamefont {Wehner},
  \citenamefont {Elkouss},\ and\ \citenamefont
  {Hanson}}]{Wehnereaam9288Quantum}%
  \BibitemOpen
  \bibfield  {author} {\bibinfo {author} {\bibfnamefont {S.}~\bibnamefont
  {Wehner}}, \bibinfo {author} {\bibfnamefont {D.}~\bibnamefont {Elkouss}}, \
  and\ \bibinfo {author} {\bibfnamefont {R.}~\bibnamefont {Hanson}},\
  }\href@noop {} {\bibfield  {journal} {\bibinfo  {journal} {Science}\ }\textbf
  {\bibinfo {volume} {362}} (\bibinfo {year} {2018})}\BibitemShut {NoStop}%
\bibitem [{\citenamefont {Caleffi}\ \emph {et~al.}(2018)\citenamefont
  {Caleffi}, \citenamefont {Cacciapuoti},\ and\ \citenamefont
  {Bianchi}}]{caleffi2018quantum}%
  \BibitemOpen
  \bibfield  {author} {\bibinfo {author} {\bibfnamefont {M.}~\bibnamefont
  {Caleffi}}, \bibinfo {author} {\bibfnamefont {A.~S.}\ \bibnamefont
  {Cacciapuoti}}, \ and\ \bibinfo {author} {\bibfnamefont {G.}~\bibnamefont
  {Bianchi}},\ }\href@noop {} {\bibfield  {journal} {\bibinfo  {journal}
  {arXiv:1805.04360}\ } (\bibinfo {year} {2018})}\BibitemShut {NoStop}%
\bibitem [{\citenamefont {Castelvecchi}(2018)}]{castelvecchi2018quantum}%
  \BibitemOpen
  \bibfield  {author} {\bibinfo {author} {\bibfnamefont {D.}~\bibnamefont
  {Castelvecchi}},\ }\href@noop {} {\bibfield  {journal} {\bibinfo  {journal}
  {Nature}\ }\textbf {\bibinfo {volume} {554}} (\bibinfo {year}
  {2018})}\BibitemShut {NoStop}%
\bibitem [{\citenamefont {Steane}(1998)}]{Steane_1998}%
  \BibitemOpen
  \bibfield  {author} {\bibinfo {author} {\bibfnamefont {A.}~\bibnamefont
  {Steane}},\ }\href@noop {} {\bibfield  {journal} {\bibinfo  {journal} {Rep.
  Pro. Phys.}\ }\textbf {\bibinfo {volume} {61}},\ \bibinfo {pages} {117}
  (\bibinfo {year} {1998})}\BibitemShut {NoStop}%
\bibitem [{\citenamefont {Gisin}\ and\ \citenamefont {Thew}(2007)}]{Gisin2007}%
  \BibitemOpen
  \bibfield  {author} {\bibinfo {author} {\bibfnamefont {N.}~\bibnamefont
  {Gisin}}\ and\ \bibinfo {author} {\bibfnamefont {R.}~\bibnamefont {Thew}},\
  }\href@noop {} {\bibfield  {journal} {\bibinfo  {journal} {Nat. photon.}\
  }\textbf {\bibinfo {volume} {1}},\ \bibinfo {pages} {165} (\bibinfo {year}
  {2007})}\BibitemShut {NoStop}%
\bibitem [{\citenamefont {Giovannetti}\ \emph {et~al.}(2006)\citenamefont
  {Giovannetti}, \citenamefont {Lloyd},\ and\ \citenamefont
  {Maccone}}]{Giovannetti2006}%
  \BibitemOpen
  \bibfield  {author} {\bibinfo {author} {\bibfnamefont {V.}~\bibnamefont
  {Giovannetti}}, \bibinfo {author} {\bibfnamefont {S.}~\bibnamefont {Lloyd}},
  \ and\ \bibinfo {author} {\bibfnamefont {L.}~\bibnamefont {Maccone}},\
  }\href@noop {} {\bibfield  {journal} {\bibinfo  {journal} {Phys. Rev. Lett.}\
  }\textbf {\bibinfo {volume} {96}},\ \bibinfo {pages} {010401} (\bibinfo
  {year} {2006})}\BibitemShut {NoStop}%
\bibitem [{\citenamefont {Bose}\ \emph {et~al.}(1998)\citenamefont {Bose},
  \citenamefont {Vedral},\ and\ \citenamefont
  {Knight}}]{bose1998multiparticle}%
  \BibitemOpen
  \bibfield  {author} {\bibinfo {author} {\bibfnamefont {S.}~\bibnamefont
  {Bose}}, \bibinfo {author} {\bibfnamefont {V.}~\bibnamefont {Vedral}}, \ and\
  \bibinfo {author} {\bibfnamefont {P.~L.}\ \bibnamefont {Knight}},\
  }\href@noop {} {\bibfield  {journal} {\bibinfo  {journal} {Phys. Rev. A}\
  }\textbf {\bibinfo {volume} {57}},\ \bibinfo {pages} {822} (\bibinfo {year}
  {1998})}\BibitemShut {NoStop}%
\bibitem [{\citenamefont {Chen}\ and\ \citenamefont
  {Lo}(2007)}]{chen2007multi}%
  \BibitemOpen
  \bibfield  {author} {\bibinfo {author} {\bibfnamefont {K.}~\bibnamefont
  {Chen}}\ and\ \bibinfo {author} {\bibfnamefont {H.-K.}\ \bibnamefont {Lo}},\
  }\href@noop {} {\bibfield  {journal} {\bibinfo  {journal} {Quantum Inf.
  Comput.}\ }\textbf {\bibinfo {volume} {7}},\ \bibinfo {pages} {689} (\bibinfo
  {year} {2007})}\BibitemShut {NoStop}%
\bibitem [{\citenamefont {Chen}\ and\ \citenamefont
  {Lo}(2005)}]{chen2005conference}%
  \BibitemOpen
  \bibfield  {author} {\bibinfo {author} {\bibfnamefont {K.}~\bibnamefont
  {Chen}}\ and\ \bibinfo {author} {\bibfnamefont {H.-K.}\ \bibnamefont {Lo}},\
  }in\ \href@noop {} {\emph {\bibinfo {booktitle} {Proceedings of the 2005 IEEE
  International Symposium on Information Theory}}}\ (\bibinfo {organization}
  {IEEE},\ \bibinfo {year} {Adelaide, Australia, 2005})\ pp.\ \bibinfo {pages}
  {1607--1611}\BibitemShut {NoStop}%
\bibitem [{\citenamefont {Fu}\ \emph {et~al.}(2015)\citenamefont {Fu},
  \citenamefont {Yin}, \citenamefont {Chen},\ and\ \citenamefont
  {Chen}}]{fu2015long}%
  \BibitemOpen
  \bibfield  {author} {\bibinfo {author} {\bibfnamefont {Y.}~\bibnamefont
  {Fu}}, \bibinfo {author} {\bibfnamefont {H.-L.}\ \bibnamefont {Yin}},
  \bibinfo {author} {\bibfnamefont {T.-Y.}\ \bibnamefont {Chen}}, \ and\
  \bibinfo {author} {\bibfnamefont {Z.-B.}\ \bibnamefont {Chen}},\ }\href@noop
  {} {\bibfield  {journal} {\bibinfo  {journal} {Phys. Rev. Lett.}\ }\textbf
  {\bibinfo {volume} {114}},\ \bibinfo {pages} {090501} (\bibinfo {year}
  {2015})}\BibitemShut {NoStop}%
\bibitem [{\citenamefont {Grasselli}\ \emph {et~al.}(2019)\citenamefont
  {Grasselli}, \citenamefont {Kampermann},\ and\ \citenamefont
  {Bru{\ss}}}]{grasselli2019conference}%
  \BibitemOpen
  \bibfield  {author} {\bibinfo {author} {\bibfnamefont {F.}~\bibnamefont
  {Grasselli}}, \bibinfo {author} {\bibfnamefont {H.}~\bibnamefont
  {Kampermann}}, \ and\ \bibinfo {author} {\bibfnamefont {D.}~\bibnamefont
  {Bru{\ss}}},\ }\href@noop {} {\bibfield  {journal} {\bibinfo  {journal} {New
  J. Phys.}\ }\textbf {\bibinfo {volume} {21}},\ \bibinfo {pages} {123002}
  (\bibinfo {year} {2019})}\BibitemShut {NoStop}%
\bibitem [{\citenamefont {Murta}\ \emph {et~al.}(2020)\citenamefont {Murta},
  \citenamefont {Grasselli}, \citenamefont {Kampermann},\ and\ \citenamefont
  {Bru{\ss}}}]{murta2020quantum}%
  \BibitemOpen
  \bibfield  {author} {\bibinfo {author} {\bibfnamefont {G.}~\bibnamefont
  {Murta}}, \bibinfo {author} {\bibfnamefont {F.}~\bibnamefont {Grasselli}},
  \bibinfo {author} {\bibfnamefont {H.}~\bibnamefont {Kampermann}}, \ and\
  \bibinfo {author} {\bibfnamefont {D.}~\bibnamefont {Bru{\ss}}},\ }\href@noop
  {} {\bibfield  {journal} {\bibinfo  {journal} {arXiv preprint
  arXiv:2003.10186}\ } (\bibinfo {year} {2020})}\BibitemShut {NoStop}%
\bibitem [{\citenamefont {Greenberger}\ \emph {et~al.}(1989)\citenamefont
  {Greenberger}, \citenamefont {Horne},\ and\ \citenamefont
  {Zeilinger}}]{Greenberger1989}%
  \BibitemOpen
  \bibfield  {author} {\bibinfo {author} {\bibfnamefont {D.~M.}\ \bibnamefont
  {Greenberger}}, \bibinfo {author} {\bibfnamefont {M.~A.}\ \bibnamefont
  {Horne}}, \ and\ \bibinfo {author} {\bibfnamefont {A.}~\bibnamefont
  {Zeilinger}},\ }in\ \href@noop {} {\emph {\bibinfo {booktitle} {Bell's
  theorem, quantum theory and conceptions of the universe}}}\ (\bibinfo
  {publisher} {Springer},\ \bibinfo {year} {1989})\ pp.\ \bibinfo {pages}
  {69--72}\BibitemShut {NoStop}%
\bibitem [{\citenamefont {Bell}(1964)}]{bell1964einstein}%
  \BibitemOpen
  \bibfield  {author} {\bibinfo {author} {\bibfnamefont {J.~S.}\ \bibnamefont
  {Bell}},\ }\href@noop {} {\bibfield  {journal} {\bibinfo  {journal} {Phys.}\
  }\textbf {\bibinfo {volume} {1}},\ \bibinfo {pages} {195} (\bibinfo {year}
  {1964})}\BibitemShut {NoStop}%
\bibitem [{\citenamefont {Brunner}\ \emph {et~al.}(2014)\citenamefont
  {Brunner}, \citenamefont {Cavalcanti}, \citenamefont {Pironio}, \citenamefont
  {Scarani},\ and\ \citenamefont {Wehner}}]{brunner2014bell}%
  \BibitemOpen
  \bibfield  {author} {\bibinfo {author} {\bibfnamefont {N.}~\bibnamefont
  {Brunner}}, \bibinfo {author} {\bibfnamefont {D.}~\bibnamefont {Cavalcanti}},
  \bibinfo {author} {\bibfnamefont {S.}~\bibnamefont {Pironio}}, \bibinfo
  {author} {\bibfnamefont {V.}~\bibnamefont {Scarani}}, \ and\ \bibinfo
  {author} {\bibfnamefont {S.}~\bibnamefont {Wehner}},\ }\href@noop {}
  {\bibfield  {journal} {\bibinfo  {journal} {Rev. Mod. Phys.}\ }\textbf
  {\bibinfo {volume} {86}},\ \bibinfo {pages} {419} (\bibinfo {year}
  {2014})}\BibitemShut {NoStop}%
\bibitem [{\citenamefont {Bourennane}\ \emph {et~al.}(2003)\citenamefont
  {Bourennane}, \citenamefont {Eibl}, \citenamefont {Gaertner}, \citenamefont
  {Kiesel}, \citenamefont {Kurtsiefer}, \citenamefont {{\.Z}ukowski},\ and\
  \citenamefont {Weinfurter}}]{bourennane2003multiphoton}%
  \BibitemOpen
  \bibfield  {author} {\bibinfo {author} {\bibfnamefont {M.}~\bibnamefont
  {Bourennane}}, \bibinfo {author} {\bibfnamefont {M.}~\bibnamefont {Eibl}},
  \bibinfo {author} {\bibfnamefont {S.}~\bibnamefont {Gaertner}}, \bibinfo
  {author} {\bibfnamefont {N.}~\bibnamefont {Kiesel}}, \bibinfo {author}
  {\bibfnamefont {C.}~\bibnamefont {Kurtsiefer}}, \bibinfo {author}
  {\bibfnamefont {M.}~\bibnamefont {{\.Z}ukowski}}, \ and\ \bibinfo {author}
  {\bibfnamefont {H.}~\bibnamefont {Weinfurter}},\ }\href@noop {} {\bibfield
  {journal} {\bibinfo  {journal} {Fortschritte der Physik: Progress of
  Physics}\ }\textbf {\bibinfo {volume} {51}},\ \bibinfo {pages} {273}
  (\bibinfo {year} {2003})}\BibitemShut {NoStop}%
\bibitem [{\citenamefont {Pan}\ \emph {et~al.}(2012)\citenamefont {Pan},
  \citenamefont {Chen}, \citenamefont {Lu}, \citenamefont {Weinfurter},
  \citenamefont {Zeilinger},\ and\ \citenamefont
  {{\.Z}ukowski}}]{pan2012multiphoton}%
  \BibitemOpen
  \bibfield  {author} {\bibinfo {author} {\bibfnamefont {J.-W.}\ \bibnamefont
  {Pan}}, \bibinfo {author} {\bibfnamefont {Z.-B.}\ \bibnamefont {Chen}},
  \bibinfo {author} {\bibfnamefont {C.-Y.}\ \bibnamefont {Lu}}, \bibinfo
  {author} {\bibfnamefont {H.}~\bibnamefont {Weinfurter}}, \bibinfo {author}
  {\bibfnamefont {A.}~\bibnamefont {Zeilinger}}, \ and\ \bibinfo {author}
  {\bibfnamefont {M.}~\bibnamefont {{\.Z}ukowski}},\ }\href@noop {} {\bibfield
  {journal} {\bibinfo  {journal} {Rev. Mod. Phys.}\ }\textbf {\bibinfo {volume}
  {84}},\ \bibinfo {pages} {777} (\bibinfo {year} {2012})}\BibitemShut
  {NoStop}%
\bibitem [{\citenamefont {Monz}\ \emph {et~al.}(2011)\citenamefont {Monz},
  \citenamefont {Schindler}, \citenamefont {Barreiro}, \citenamefont {Chwalla},
  \citenamefont {Nigg}, \citenamefont {Coish}, \citenamefont {Harlander},
  \citenamefont {H{\"a}nsel}, \citenamefont {Hennrich},\ and\ \citenamefont
  {Blatt}}]{monz201114}%
  \BibitemOpen
  \bibfield  {author} {\bibinfo {author} {\bibfnamefont {T.}~\bibnamefont
  {Monz}}, \bibinfo {author} {\bibfnamefont {P.}~\bibnamefont {Schindler}},
  \bibinfo {author} {\bibfnamefont {J.~T.}\ \bibnamefont {Barreiro}}, \bibinfo
  {author} {\bibfnamefont {M.}~\bibnamefont {Chwalla}}, \bibinfo {author}
  {\bibfnamefont {D.}~\bibnamefont {Nigg}}, \bibinfo {author} {\bibfnamefont
  {W.~A.}\ \bibnamefont {Coish}}, \bibinfo {author} {\bibfnamefont
  {M.}~\bibnamefont {Harlander}}, \bibinfo {author} {\bibfnamefont
  {W.}~\bibnamefont {H{\"a}nsel}}, \bibinfo {author} {\bibfnamefont
  {M.}~\bibnamefont {Hennrich}}, \ and\ \bibinfo {author} {\bibfnamefont
  {R.}~\bibnamefont {Blatt}},\ }\href@noop {} {\bibfield  {journal} {\bibinfo
  {journal} {Phys. Rev. Lett.}\ }\textbf {\bibinfo {volume} {106}},\ \bibinfo
  {pages} {130506} (\bibinfo {year} {2011})}\BibitemShut {NoStop}%
\bibitem [{\citenamefont {Wang}\ \emph {et~al.}(2016)\citenamefont {Wang},
  \citenamefont {Chen}, \citenamefont {Li}, \citenamefont {Huang},
  \citenamefont {Liu}, \citenamefont {Chen}, \citenamefont {Luo}, \citenamefont
  {Su}, \citenamefont {Wu}, \citenamefont {Li} \emph
  {et~al.}}]{wang2016experimental}%
  \BibitemOpen
  \bibfield  {author} {\bibinfo {author} {\bibfnamefont {X.-L.}\ \bibnamefont
  {Wang}}, \bibinfo {author} {\bibfnamefont {L.-K.}\ \bibnamefont {Chen}},
  \bibinfo {author} {\bibfnamefont {W.}~\bibnamefont {Li}}, \bibinfo {author}
  {\bibfnamefont {H.-L.}\ \bibnamefont {Huang}}, \bibinfo {author}
  {\bibfnamefont {C.}~\bibnamefont {Liu}}, \bibinfo {author} {\bibfnamefont
  {C.}~\bibnamefont {Chen}}, \bibinfo {author} {\bibfnamefont {Y.-H.}\
  \bibnamefont {Luo}}, \bibinfo {author} {\bibfnamefont {Z.-E.}\ \bibnamefont
  {Su}}, \bibinfo {author} {\bibfnamefont {D.}~\bibnamefont {Wu}}, \bibinfo
  {author} {\bibfnamefont {Z.-D.}\ \bibnamefont {Li}},  \emph {et~al.},\
  }\href@noop {} {\bibfield  {journal} {\bibinfo  {journal} {Phys. Rev. Lett.}\
  }\textbf {\bibinfo {volume} {117}},\ \bibinfo {pages} {210502} (\bibinfo
  {year} {2016})}\BibitemShut {NoStop}%
\bibitem [{\citenamefont {Song}\ \emph {et~al.}(2017)\citenamefont {Song},
  \citenamefont {Xu}, \citenamefont {Liu}, \citenamefont {Yang}, \citenamefont
  {Zheng}, \citenamefont {Deng}, \citenamefont {Xie}, \citenamefont {Huang},
  \citenamefont {Guo}, \citenamefont {Zhang} \emph {et~al.}}]{song201710}%
  \BibitemOpen
  \bibfield  {author} {\bibinfo {author} {\bibfnamefont {C.}~\bibnamefont
  {Song}}, \bibinfo {author} {\bibfnamefont {K.}~\bibnamefont {Xu}}, \bibinfo
  {author} {\bibfnamefont {W.}~\bibnamefont {Liu}}, \bibinfo {author}
  {\bibfnamefont {C.-P.}\ \bibnamefont {Yang}}, \bibinfo {author}
  {\bibfnamefont {S.-B.}\ \bibnamefont {Zheng}}, \bibinfo {author}
  {\bibfnamefont {H.}~\bibnamefont {Deng}}, \bibinfo {author} {\bibfnamefont
  {Q.}~\bibnamefont {Xie}}, \bibinfo {author} {\bibfnamefont {K.}~\bibnamefont
  {Huang}}, \bibinfo {author} {\bibfnamefont {Q.}~\bibnamefont {Guo}}, \bibinfo
  {author} {\bibfnamefont {L.}~\bibnamefont {Zhang}},  \emph {et~al.},\
  }\href@noop {} {\bibfield  {journal} {\bibinfo  {journal} {Phys. Rev. Lett.}\
  }\textbf {\bibinfo {volume} {119}},\ \bibinfo {pages} {180511} (\bibinfo
  {year} {2017})}\BibitemShut {NoStop}%
\bibitem [{\citenamefont {Chen}\ \emph {et~al.}(2017)\citenamefont {Chen},
  \citenamefont {Li}, \citenamefont {Yao}, \citenamefont {Huang}, \citenamefont
  {Li}, \citenamefont {Lu}, \citenamefont {Yuan}, \citenamefont {Zhang},
  \citenamefont {Jiang}, \citenamefont {Peng} \emph
  {et~al.}}]{chen2017observation}%
  \BibitemOpen
  \bibfield  {author} {\bibinfo {author} {\bibfnamefont {L.-K.}\ \bibnamefont
  {Chen}}, \bibinfo {author} {\bibfnamefont {Z.-D.}\ \bibnamefont {Li}},
  \bibinfo {author} {\bibfnamefont {X.-C.}\ \bibnamefont {Yao}}, \bibinfo
  {author} {\bibfnamefont {M.}~\bibnamefont {Huang}}, \bibinfo {author}
  {\bibfnamefont {W.}~\bibnamefont {Li}}, \bibinfo {author} {\bibfnamefont
  {H.}~\bibnamefont {Lu}}, \bibinfo {author} {\bibfnamefont {X.}~\bibnamefont
  {Yuan}}, \bibinfo {author} {\bibfnamefont {Y.-B.}\ \bibnamefont {Zhang}},
  \bibinfo {author} {\bibfnamefont {X.}~\bibnamefont {Jiang}}, \bibinfo
  {author} {\bibfnamefont {C.-Z.}\ \bibnamefont {Peng}},  \emph {et~al.},\
  }\href@noop {} {\bibfield  {journal} {\bibinfo  {journal} {Optica}\ }\textbf
  {\bibinfo {volume} {4}},\ \bibinfo {pages} {77} (\bibinfo {year}
  {2017})}\BibitemShut {NoStop}%
\bibitem [{\citenamefont {Wang}\ \emph
  {et~al.}(2018{\natexlab{a}})\citenamefont {Wang}, \citenamefont {Luo},
  \citenamefont {Huang}, \citenamefont {Chen}, \citenamefont {Su},
  \citenamefont {Liu}, \citenamefont {Chen}, \citenamefont {Li}, \citenamefont
  {Fang}, \citenamefont {Jiang} \emph {et~al.}}]{wang201818}%
  \BibitemOpen
  \bibfield  {author} {\bibinfo {author} {\bibfnamefont {X.-L.}\ \bibnamefont
  {Wang}}, \bibinfo {author} {\bibfnamefont {Y.-H.}\ \bibnamefont {Luo}},
  \bibinfo {author} {\bibfnamefont {H.-L.}\ \bibnamefont {Huang}}, \bibinfo
  {author} {\bibfnamefont {M.-C.}\ \bibnamefont {Chen}}, \bibinfo {author}
  {\bibfnamefont {Z.-E.}\ \bibnamefont {Su}}, \bibinfo {author} {\bibfnamefont
  {C.}~\bibnamefont {Liu}}, \bibinfo {author} {\bibfnamefont {C.}~\bibnamefont
  {Chen}}, \bibinfo {author} {\bibfnamefont {W.}~\bibnamefont {Li}}, \bibinfo
  {author} {\bibfnamefont {Y.-Q.}\ \bibnamefont {Fang}}, \bibinfo {author}
  {\bibfnamefont {X.}~\bibnamefont {Jiang}},  \emph {et~al.},\ }\href@noop {}
  {\bibfield  {journal} {\bibinfo  {journal} {Phys. Rev. Lett.}\ }\textbf
  {\bibinfo {volume} {120}},\ \bibinfo {pages} {260502} (\bibinfo {year}
  {2018}{\natexlab{a}})}\BibitemShut {NoStop}%
\bibitem [{\citenamefont {Qian}\ \emph {et~al.}(2005)\citenamefont {Qian},
  \citenamefont {Feng},\ and\ \citenamefont {Gong}}]{qian2005universal}%
  \BibitemOpen
  \bibfield  {author} {\bibinfo {author} {\bibfnamefont {J.}~\bibnamefont
  {Qian}}, \bibinfo {author} {\bibfnamefont {X.-L.}\ \bibnamefont {Feng}}, \
  and\ \bibinfo {author} {\bibfnamefont {S.-Q.}\ \bibnamefont {Gong}},\
  }\href@noop {} {\bibfield  {journal} {\bibinfo  {journal} {Phys. Rev. A}\
  }\textbf {\bibinfo {volume} {72}},\ \bibinfo {pages} {052308} (\bibinfo
  {year} {2005})}\BibitemShut {NoStop}%
\bibitem [{\citenamefont {Lo}\ \emph {et~al.}(2012)\citenamefont {Lo},
  \citenamefont {Curty},\ and\ \citenamefont {Qi}}]{lo2012measurement}%
  \BibitemOpen
  \bibfield  {author} {\bibinfo {author} {\bibfnamefont {H.-K.}\ \bibnamefont
  {Lo}}, \bibinfo {author} {\bibfnamefont {M.}~\bibnamefont {Curty}}, \ and\
  \bibinfo {author} {\bibfnamefont {B.}~\bibnamefont {Qi}},\ }\href@noop {}
  {\bibfield  {journal} {\bibinfo  {journal} {Phys. Rev. Lett.}\ }\textbf
  {\bibinfo {volume} {108}},\ \bibinfo {pages} {130503} (\bibinfo {year}
  {2012})}\BibitemShut {NoStop}%
\bibitem [{\citenamefont {Braunstein}\ and\ \citenamefont
  {Pirandola}(2012)}]{braunstein2012side}%
  \BibitemOpen
  \bibfield  {author} {\bibinfo {author} {\bibfnamefont {S.~L.}\ \bibnamefont
  {Braunstein}}\ and\ \bibinfo {author} {\bibfnamefont {S.}~\bibnamefont
  {Pirandola}},\ }\href@noop {} {\bibfield  {journal} {\bibinfo  {journal}
  {Phys. Rev. Lett.}\ }\textbf {\bibinfo {volume} {108}},\ \bibinfo {pages}
  {130502} (\bibinfo {year} {2012})}\BibitemShut {NoStop}%
\bibitem [{\citenamefont {Lo}\ \emph {et~al.}(2005)\citenamefont {Lo},
  \citenamefont {Ma},\ and\ \citenamefont {Chen}}]{lo2005decoy}%
  \BibitemOpen
  \bibfield  {author} {\bibinfo {author} {\bibfnamefont {H.-K.}\ \bibnamefont
  {Lo}}, \bibinfo {author} {\bibfnamefont {X.}~\bibnamefont {Ma}}, \ and\
  \bibinfo {author} {\bibfnamefont {K.}~\bibnamefont {Chen}},\ }\href@noop {}
  {\bibfield  {journal} {\bibinfo  {journal} {Phys. Rev. Lett.}\ }\textbf
  {\bibinfo {volume} {94}},\ \bibinfo {pages} {230504} (\bibinfo {year}
  {2005})}\BibitemShut {NoStop}%
\bibitem [{\citenamefont {D{\"u}r}\ \emph {et~al.}(2000)\citenamefont
  {D{\"u}r}, \citenamefont {Vidal},\ and\ \citenamefont
  {Cirac}}]{dur2000three}%
  \BibitemOpen
  \bibfield  {author} {\bibinfo {author} {\bibfnamefont {W.}~\bibnamefont
  {D{\"u}r}}, \bibinfo {author} {\bibfnamefont {G.}~\bibnamefont {Vidal}}, \
  and\ \bibinfo {author} {\bibfnamefont {J.~I.}\ \bibnamefont {Cirac}},\
  }\href@noop {} {\bibfield  {journal} {\bibinfo  {journal} {Phys. Rev. A}\
  }\textbf {\bibinfo {volume} {62}},\ \bibinfo {pages} {062314} (\bibinfo
  {year} {2000})}\BibitemShut {NoStop}%
\bibitem [{\citenamefont {Lucamarini}\ \emph {et~al.}(2018)\citenamefont
  {Lucamarini}, \citenamefont {Yuan}, \citenamefont {Dynes},\ and\
  \citenamefont {Shields}}]{lucamarini2018overcoming}%
  \BibitemOpen
  \bibfield  {author} {\bibinfo {author} {\bibfnamefont {M.}~\bibnamefont
  {Lucamarini}}, \bibinfo {author} {\bibfnamefont {Z.}~\bibnamefont {Yuan}},
  \bibinfo {author} {\bibfnamefont {J.}~\bibnamefont {Dynes}}, \ and\ \bibinfo
  {author} {\bibfnamefont {A.}~\bibnamefont {Shields}},\ }\href@noop {}
  {\bibfield  {journal} {\bibinfo  {journal} {Nature}\ }\textbf {\bibinfo
  {volume} {557}},\ \bibinfo {pages} {400} (\bibinfo {year}
  {2018})}\BibitemShut {NoStop}%
\bibitem [{\citenamefont {Ma}\ \emph {et~al.}(2018)\citenamefont {Ma},
  \citenamefont {Zeng},\ and\ \citenamefont {Zhou}}]{ma2018phase}%
  \BibitemOpen
  \bibfield  {author} {\bibinfo {author} {\bibfnamefont {X.}~\bibnamefont
  {Ma}}, \bibinfo {author} {\bibfnamefont {P.}~\bibnamefont {Zeng}}, \ and\
  \bibinfo {author} {\bibfnamefont {H.}~\bibnamefont {Zhou}},\ }\href@noop {}
  {\bibfield  {journal} {\bibinfo  {journal} {Phys. Rev. X}\ }\textbf {\bibinfo
  {volume} {8}},\ \bibinfo {pages} {031043} (\bibinfo {year}
  {2018})}\BibitemShut {NoStop}%
\bibitem [{\citenamefont {Tamaki}\ \emph {et~al.}(2018)\citenamefont {Tamaki},
  \citenamefont {Lo}, \citenamefont {Wang},\ and\ \citenamefont
  {Lucamarini}}]{tamaki2018information}%
  \BibitemOpen
  \bibfield  {author} {\bibinfo {author} {\bibfnamefont {K.}~\bibnamefont
  {Tamaki}}, \bibinfo {author} {\bibfnamefont {H.-K.}\ \bibnamefont {Lo}},
  \bibinfo {author} {\bibfnamefont {W.}~\bibnamefont {Wang}}, \ and\ \bibinfo
  {author} {\bibfnamefont {M.}~\bibnamefont {Lucamarini}},\ }\href@noop {}
  {\bibfield  {journal} {\bibinfo  {journal} {arXiv:1805.05511}\ } (\bibinfo
  {year} {2018})}\BibitemShut {NoStop}%
\bibitem [{\citenamefont {Wang}\ \emph
  {et~al.}(2018{\natexlab{b}})\citenamefont {Wang}, \citenamefont {Yu},\ and\
  \citenamefont {Hu}}]{wang2018twin}%
  \BibitemOpen
  \bibfield  {author} {\bibinfo {author} {\bibfnamefont {X.-B.}\ \bibnamefont
  {Wang}}, \bibinfo {author} {\bibfnamefont {Z.-W.}\ \bibnamefont {Yu}}, \ and\
  \bibinfo {author} {\bibfnamefont {X.-L.}\ \bibnamefont {Hu}},\ }\href@noop {}
  {\bibfield  {journal} {\bibinfo  {journal} {Phys. Rev. A}\ }\textbf {\bibinfo
  {volume} {98}},\ \bibinfo {pages} {062323} (\bibinfo {year}
  {2018}{\natexlab{b}})}\BibitemShut {NoStop}%
\bibitem [{\citenamefont {Cui}\ \emph {et~al.}(2019)\citenamefont {Cui},
  \citenamefont {Yin}, \citenamefont {Wang}, \citenamefont {Chen},
  \citenamefont {Wang}, \citenamefont {Guo},\ and\ \citenamefont
  {Han}}]{Cui2019Twin-Field}%
  \BibitemOpen
  \bibfield  {author} {\bibinfo {author} {\bibfnamefont {C.}~\bibnamefont
  {Cui}}, \bibinfo {author} {\bibfnamefont {Z.-Q.}\ \bibnamefont {Yin}},
  \bibinfo {author} {\bibfnamefont {R.}~\bibnamefont {Wang}}, \bibinfo {author}
  {\bibfnamefont {W.}~\bibnamefont {Chen}}, \bibinfo {author} {\bibfnamefont
  {S.}~\bibnamefont {Wang}}, \bibinfo {author} {\bibfnamefont {G.-C.}\
  \bibnamefont {Guo}}, \ and\ \bibinfo {author} {\bibfnamefont {Z.-F.}\
  \bibnamefont {Han}},\ }\href@noop {} {\bibfield  {journal} {\bibinfo
  {journal} {Phys. Rev. Applied}\ }\textbf {\bibinfo {volume} {11}},\ \bibinfo
  {pages} {034053} (\bibinfo {year} {2019})}\BibitemShut {NoStop}%
\bibitem [{\citenamefont {Lin}\ and\ \citenamefont
  {L{\"u}tkenhaus}(2018)}]{lin2018simple}%
  \BibitemOpen
  \bibfield  {author} {\bibinfo {author} {\bibfnamefont {J.}~\bibnamefont
  {Lin}}\ and\ \bibinfo {author} {\bibfnamefont {N.}~\bibnamefont
  {L{\"u}tkenhaus}},\ }\href@noop {} {\bibfield  {journal} {\bibinfo  {journal}
  {Phys. Rev. A}\ }\textbf {\bibinfo {volume} {98}},\ \bibinfo {pages} {042332}
  (\bibinfo {year} {2018})}\BibitemShut {NoStop}%
\bibitem [{\citenamefont {Yu}\ \emph {et~al.}(2019)\citenamefont {Yu},
  \citenamefont {Hu}, \citenamefont {Jiang}, \citenamefont {Xu},\ and\
  \citenamefont {Wang}}]{yu2019sending}%
  \BibitemOpen
  \bibfield  {author} {\bibinfo {author} {\bibfnamefont {Z.-W.}\ \bibnamefont
  {Yu}}, \bibinfo {author} {\bibfnamefont {X.-L.}\ \bibnamefont {Hu}}, \bibinfo
  {author} {\bibfnamefont {C.}~\bibnamefont {Jiang}}, \bibinfo {author}
  {\bibfnamefont {H.}~\bibnamefont {Xu}}, \ and\ \bibinfo {author}
  {\bibfnamefont {X.-B.}\ \bibnamefont {Wang}},\ }\href@noop {} {\bibfield
  {journal} {\bibinfo  {journal} {Sci. Rep.}\ }\textbf {\bibinfo {volume}
  {9}},\ \bibinfo {pages} {3080} (\bibinfo {year} {2019})}\BibitemShut
  {NoStop}%
\bibitem [{\citenamefont {Curty}\ \emph {et~al.}(2019)\citenamefont {Curty},
  \citenamefont {Azuma},\ and\ \citenamefont {Lo}}]{curty2019simple}%
  \BibitemOpen
  \bibfield  {author} {\bibinfo {author} {\bibfnamefont {M.}~\bibnamefont
  {Curty}}, \bibinfo {author} {\bibfnamefont {K.}~\bibnamefont {Azuma}}, \ and\
  \bibinfo {author} {\bibfnamefont {H.-K.}\ \bibnamefont {Lo}},\ }\href@noop {}
  {\bibfield  {journal} {\bibinfo  {journal} {Npj Quantum Inf.}\ }\textbf
  {\bibinfo {volume} {5}},\ \bibinfo {pages} {64} (\bibinfo {year}
  {2019})}\BibitemShut {NoStop}%
\bibitem [{\citenamefont {Maeda}\ \emph {et~al.}(2019)\citenamefont {Maeda},
  \citenamefont {Sasaki},\ and\ \citenamefont
  {Koashi}}]{maeda2019repeaterless}%
  \BibitemOpen
  \bibfield  {author} {\bibinfo {author} {\bibfnamefont {K.}~\bibnamefont
  {Maeda}}, \bibinfo {author} {\bibfnamefont {T.}~\bibnamefont {Sasaki}}, \
  and\ \bibinfo {author} {\bibfnamefont {M.}~\bibnamefont {Koashi}},\
  }\href@noop {} {\bibfield  {journal} {\bibinfo  {journal} {Nat. commun.}\
  }\textbf {\bibinfo {volume} {10}},\ \bibinfo {pages} {1} (\bibinfo {year}
  {2019})}\BibitemShut {NoStop}%
\bibitem [{\citenamefont {Grasselli}\ and\ \citenamefont
  {Curty}(2019)}]{grasselli2019practical}%
  \BibitemOpen
  \bibfield  {author} {\bibinfo {author} {\bibfnamefont {F.}~\bibnamefont
  {Grasselli}}\ and\ \bibinfo {author} {\bibfnamefont {M.}~\bibnamefont
  {Curty}},\ }\href@noop {} {\bibfield  {journal} {\bibinfo  {journal} {New J.
  Phys.}\ }\textbf {\bibinfo {volume} {21}},\ \bibinfo {pages} {073001}
  (\bibinfo {year} {2019})}\BibitemShut {NoStop}%
\bibitem [{\citenamefont {Pirandola}\ \emph {et~al.}(2017)\citenamefont
  {Pirandola}, \citenamefont {Laurenza}, \citenamefont {Ottaviani},\ and\
  \citenamefont {Banchi}}]{pirandola2017fundamental}%
  \BibitemOpen
  \bibfield  {author} {\bibinfo {author} {\bibfnamefont {S.}~\bibnamefont
  {Pirandola}}, \bibinfo {author} {\bibfnamefont {R.}~\bibnamefont {Laurenza}},
  \bibinfo {author} {\bibfnamefont {C.}~\bibnamefont {Ottaviani}}, \ and\
  \bibinfo {author} {\bibfnamefont {L.}~\bibnamefont {Banchi}},\ }\href@noop {}
  {\bibfield  {journal} {\bibinfo  {journal} {Nat. commun.}\ }\textbf {\bibinfo
  {volume} {8}},\ \bibinfo {pages} {15043} (\bibinfo {year}
  {2017})}\BibitemShut {NoStop}%
\bibitem [{\citenamefont {Wang}\ \emph {et~al.}(2019)\citenamefont {Wang},
  \citenamefont {He}, \citenamefont {Yin}, \citenamefont {Lu}, \citenamefont
  {Cui}, \citenamefont {Chen}, \citenamefont {Zhou}, \citenamefont {Guo},\ and\
  \citenamefont {Han}}]{wang2019beating}%
  \BibitemOpen
  \bibfield  {author} {\bibinfo {author} {\bibfnamefont {S.}~\bibnamefont
  {Wang}}, \bibinfo {author} {\bibfnamefont {D.-Y.}\ \bibnamefont {He}},
  \bibinfo {author} {\bibfnamefont {Z.-Q.}\ \bibnamefont {Yin}}, \bibinfo
  {author} {\bibfnamefont {F.-Y.}\ \bibnamefont {Lu}}, \bibinfo {author}
  {\bibfnamefont {C.-H.}\ \bibnamefont {Cui}}, \bibinfo {author} {\bibfnamefont
  {W.}~\bibnamefont {Chen}}, \bibinfo {author} {\bibfnamefont {Z.}~\bibnamefont
  {Zhou}}, \bibinfo {author} {\bibfnamefont {G.-C.}\ \bibnamefont {Guo}}, \
  and\ \bibinfo {author} {\bibfnamefont {Z.-F.}\ \bibnamefont {Han}},\
  }\href@noop {} {\bibfield  {journal} {\bibinfo  {journal} {Phys. Rev. X}\
  }\textbf {\bibinfo {volume} {9}},\ \bibinfo {pages} {021046} (\bibinfo {year}
  {2019})}\BibitemShut {NoStop}%
\bibitem [{\citenamefont {Minder}\ \emph {et~al.}(2019)\citenamefont {Minder},
  \citenamefont {Pittaluga}, \citenamefont {Roberts}, \citenamefont
  {Lucamarini}, \citenamefont {Dynes}, \citenamefont {Yuan},\ and\
  \citenamefont {Shields}}]{minder2019experimental}%
  \BibitemOpen
  \bibfield  {author} {\bibinfo {author} {\bibfnamefont {M.}~\bibnamefont
  {Minder}}, \bibinfo {author} {\bibfnamefont {M.}~\bibnamefont {Pittaluga}},
  \bibinfo {author} {\bibfnamefont {G.}~\bibnamefont {Roberts}}, \bibinfo
  {author} {\bibfnamefont {M.}~\bibnamefont {Lucamarini}}, \bibinfo {author}
  {\bibfnamefont {J.}~\bibnamefont {Dynes}}, \bibinfo {author} {\bibfnamefont
  {Z.}~\bibnamefont {Yuan}}, \ and\ \bibinfo {author} {\bibfnamefont
  {A.}~\bibnamefont {Shields}},\ }\href@noop {} {\bibfield  {journal} {\bibinfo
   {journal} {Nat. Photon.}\ }\textbf {\bibinfo {volume} {13}},\ \bibinfo
  {pages} {334} (\bibinfo {year} {2019})}\BibitemShut {NoStop}%
\bibitem [{\citenamefont {Liu}\ \emph {et~al.}(2019)\citenamefont {Liu},
  \citenamefont {Yu}, \citenamefont {Zhang}, \citenamefont {Guan},
  \citenamefont {Chen}, \citenamefont {Zhang}, \citenamefont {Hu},
  \citenamefont {Li}, \citenamefont {Jiang}, \citenamefont {Lin} \emph
  {et~al.}}]{liu2019experimental}%
  \BibitemOpen
  \bibfield  {author} {\bibinfo {author} {\bibfnamefont {Y.}~\bibnamefont
  {Liu}}, \bibinfo {author} {\bibfnamefont {Z.-W.}\ \bibnamefont {Yu}},
  \bibinfo {author} {\bibfnamefont {W.}~\bibnamefont {Zhang}}, \bibinfo
  {author} {\bibfnamefont {J.-Y.}\ \bibnamefont {Guan}}, \bibinfo {author}
  {\bibfnamefont {J.-P.}\ \bibnamefont {Chen}}, \bibinfo {author}
  {\bibfnamefont {C.}~\bibnamefont {Zhang}}, \bibinfo {author} {\bibfnamefont
  {X.-L.}\ \bibnamefont {Hu}}, \bibinfo {author} {\bibfnamefont
  {H.}~\bibnamefont {Li}}, \bibinfo {author} {\bibfnamefont {C.}~\bibnamefont
  {Jiang}}, \bibinfo {author} {\bibfnamefont {J.}~\bibnamefont {Lin}},  \emph
  {et~al.},\ }\href@noop {} {\bibfield  {journal} {\bibinfo  {journal} {Phys.
  Rev. Lett.}\ }\textbf {\bibinfo {volume} {123}},\ \bibinfo {pages} {100505}
  (\bibinfo {year} {2019})}\BibitemShut {NoStop}%
\bibitem [{\citenamefont {Zhong}\ \emph {et~al.}(2019)\citenamefont {Zhong},
  \citenamefont {Hu}, \citenamefont {Curty}, \citenamefont {Qian},\ and\
  \citenamefont {Lo}}]{zhong2019proof}%
  \BibitemOpen
  \bibfield  {author} {\bibinfo {author} {\bibfnamefont {X.}~\bibnamefont
  {Zhong}}, \bibinfo {author} {\bibfnamefont {J.}~\bibnamefont {Hu}}, \bibinfo
  {author} {\bibfnamefont {M.}~\bibnamefont {Curty}}, \bibinfo {author}
  {\bibfnamefont {L.}~\bibnamefont {Qian}}, \ and\ \bibinfo {author}
  {\bibfnamefont {H.-K.}\ \bibnamefont {Lo}},\ }\href@noop {} {\bibfield
  {journal} {\bibinfo  {journal} {Phys. Rev. Lett.}\ }\textbf {\bibinfo
  {volume} {123}},\ \bibinfo {pages} {100506} (\bibinfo {year}
  {2019})}\BibitemShut {NoStop}%
\bibitem [{\citenamefont {Fang}\ \emph {et~al.}(2020)\citenamefont {Fang},
  \citenamefont {Zeng}, \citenamefont {Liu}, \citenamefont {Zou}, \citenamefont
  {Wu}, \citenamefont {Tang}, \citenamefont {Sheng}, \citenamefont {Xiang},
  \citenamefont {Zhang}, \citenamefont {Li} \emph
  {et~al.}}]{fang2020implementation}%
  \BibitemOpen
  \bibfield  {author} {\bibinfo {author} {\bibfnamefont {X.-T.}\ \bibnamefont
  {Fang}}, \bibinfo {author} {\bibfnamefont {P.}~\bibnamefont {Zeng}}, \bibinfo
  {author} {\bibfnamefont {H.}~\bibnamefont {Liu}}, \bibinfo {author}
  {\bibfnamefont {M.}~\bibnamefont {Zou}}, \bibinfo {author} {\bibfnamefont
  {W.}~\bibnamefont {Wu}}, \bibinfo {author} {\bibfnamefont {Y.-L.}\
  \bibnamefont {Tang}}, \bibinfo {author} {\bibfnamefont {Y.-J.}\ \bibnamefont
  {Sheng}}, \bibinfo {author} {\bibfnamefont {Y.}~\bibnamefont {Xiang}},
  \bibinfo {author} {\bibfnamefont {W.}~\bibnamefont {Zhang}}, \bibinfo
  {author} {\bibfnamefont {H.}~\bibnamefont {Li}},  \emph {et~al.},\
  }\href@noop {} {\bibfield  {journal} {\bibinfo  {journal} {Nat. Photon.}\ ,\
  \bibinfo {pages} {1}} (\bibinfo {year} {2020})}\BibitemShut {NoStop}%
\bibitem [{\citenamefont {Maneva}\ and\ \citenamefont
  {Smolin}(2002)}]{maneva2002improved}%
  \BibitemOpen
  \bibfield  {author} {\bibinfo {author} {\bibfnamefont {E.~N.}\ \bibnamefont
  {Maneva}}\ and\ \bibinfo {author} {\bibfnamefont {J.~A.}\ \bibnamefont
  {Smolin}},\ }\href@noop {} {\bibfield  {journal} {\bibinfo  {journal}
  {Contemp. Math.}\ }\textbf {\bibinfo {volume} {305}},\ \bibinfo {pages} {203}
  (\bibinfo {year} {2002})}\BibitemShut {NoStop}%
\bibitem [{\citenamefont {Pan}\ and\ \citenamefont
  {Zeilinger}(1998)}]{pan1998greenberger}%
  \BibitemOpen
  \bibfield  {author} {\bibinfo {author} {\bibfnamefont {J.-W.}\ \bibnamefont
  {Pan}}\ and\ \bibinfo {author} {\bibfnamefont {A.}~\bibnamefont
  {Zeilinger}},\ }\href@noop {} {\bibfield  {journal} {\bibinfo  {journal}
  {Phys. Rev. A}\ }\textbf {\bibinfo {volume} {57}},\ \bibinfo {pages} {2208}
  (\bibinfo {year} {1998})}\BibitemShut {NoStop}%
\bibitem [{\citenamefont {Lo}\ and\ \citenamefont
  {Chau}(1999)}]{lo1999unconditional}%
  \BibitemOpen
  \bibfield  {author} {\bibinfo {author} {\bibfnamefont {H.-K.}\ \bibnamefont
  {Lo}}\ and\ \bibinfo {author} {\bibfnamefont {H.~F.}\ \bibnamefont {Chau}},\
  }\href@noop {} {\bibfield  {journal} {\bibinfo  {journal} {science}\ }\textbf
  {\bibinfo {volume} {283}},\ \bibinfo {pages} {2050} (\bibinfo {year}
  {1999})}\BibitemShut {NoStop}%
\bibitem [{\citenamefont {Shor}\ and\ \citenamefont
  {Preskill}(2000)}]{shor2000simple}%
  \BibitemOpen
  \bibfield  {author} {\bibinfo {author} {\bibfnamefont {P.~W.}\ \bibnamefont
  {Shor}}\ and\ \bibinfo {author} {\bibfnamefont {J.}~\bibnamefont
  {Preskill}},\ }\href@noop {} {\bibfield  {journal} {\bibinfo  {journal}
  {Phys. Rev. Lett.}\ }\textbf {\bibinfo {volume} {85}},\ \bibinfo {pages}
  {441} (\bibinfo {year} {2000})}\BibitemShut {NoStop}%
\bibitem [{\citenamefont {Terhal}(2004)}]{terhal2004entanglement}%
  \BibitemOpen
  \bibfield  {author} {\bibinfo {author} {\bibfnamefont {B.~M.}\ \bibnamefont
  {Terhal}},\ }\href@noop {} {\bibfield  {journal} {\bibinfo  {journal} {IBM J.
  Res. Dev.}\ }\textbf {\bibinfo {volume} {48}},\ \bibinfo {pages} {71}
  (\bibinfo {year} {2004})}\BibitemShut {NoStop}%
\bibitem [{\citenamefont {Koashi}\ and\ \citenamefont
  {Winter}(2004)}]{Koashi2004Monogamy}%
  \BibitemOpen
  \bibfield  {author} {\bibinfo {author} {\bibfnamefont {M.}~\bibnamefont
  {Koashi}}\ and\ \bibinfo {author} {\bibfnamefont {A.}~\bibnamefont
  {Winter}},\ }\href@noop {} {\bibfield  {journal} {\bibinfo  {journal} {Phys.
  Rev. A}\ }\textbf {\bibinfo {volume} {69}},\ \bibinfo {pages} {022309}
  (\bibinfo {year} {2004})}\BibitemShut {NoStop}%
\bibitem [{\citenamefont {Yin}\ \emph {et~al.}(2016)\citenamefont {Yin},
  \citenamefont {Chen}, \citenamefont {Yu}, \citenamefont {Liu}, \citenamefont
  {You}, \citenamefont {Zhou}, \citenamefont {Chen}, \citenamefont {Mao},
  \citenamefont {Huang}, \citenamefont {Zhang} \emph
  {et~al.}}]{yin2016measurement}%
  \BibitemOpen
  \bibfield  {author} {\bibinfo {author} {\bibfnamefont {H.-L.}\ \bibnamefont
  {Yin}}, \bibinfo {author} {\bibfnamefont {T.-Y.}\ \bibnamefont {Chen}},
  \bibinfo {author} {\bibfnamefont {Z.-W.}\ \bibnamefont {Yu}}, \bibinfo
  {author} {\bibfnamefont {H.}~\bibnamefont {Liu}}, \bibinfo {author}
  {\bibfnamefont {L.-X.}\ \bibnamefont {You}}, \bibinfo {author} {\bibfnamefont
  {Y.-H.}\ \bibnamefont {Zhou}}, \bibinfo {author} {\bibfnamefont {S.-J.}\
  \bibnamefont {Chen}}, \bibinfo {author} {\bibfnamefont {Y.}~\bibnamefont
  {Mao}}, \bibinfo {author} {\bibfnamefont {M.-Q.}\ \bibnamefont {Huang}},
  \bibinfo {author} {\bibfnamefont {W.-J.}\ \bibnamefont {Zhang}},  \emph
  {et~al.},\ }\href@noop {} {\bibfield  {journal} {\bibinfo  {journal} {Phys.
  Rev. Lett.}\ }\textbf {\bibinfo {volume} {117}},\ \bibinfo {pages} {190501}
  (\bibinfo {year} {2016})}\BibitemShut {NoStop}%
\bibitem [{\citenamefont {Zeng}\ \emph {et~al.}(2020)\citenamefont {Zeng},
  \citenamefont {Wu},\ and\ \citenamefont {Ma}}]{zeng2019symmetry}%
  \BibitemOpen
  \bibfield  {author} {\bibinfo {author} {\bibfnamefont {P.}~\bibnamefont
  {Zeng}}, \bibinfo {author} {\bibfnamefont {W.}~\bibnamefont {Wu}}, \ and\
  \bibinfo {author} {\bibfnamefont {X.}~\bibnamefont {Ma}},\ }\href@noop {}
  {\bibfield  {journal} {\bibinfo  {journal} {Phys. Rev. Applied}\ }\textbf
  {\bibinfo {volume} {13}},\ \bibinfo {pages} {064013} (\bibinfo {year}
  {2020})}\BibitemShut {NoStop}%
\bibitem [{\citenamefont {Xu}\ \emph {et~al.}(2013)\citenamefont {Xu},
  \citenamefont {Curty}, \citenamefont {Qi},\ and\ \citenamefont
  {Lo}}]{Xu2013Practical}%
  \BibitemOpen
  \bibfield  {author} {\bibinfo {author} {\bibfnamefont {F.}~\bibnamefont
  {Xu}}, \bibinfo {author} {\bibfnamefont {M.}~\bibnamefont {Curty}}, \bibinfo
  {author} {\bibfnamefont {B.}~\bibnamefont {Qi}}, \ and\ \bibinfo {author}
  {\bibfnamefont {H.~K.}\ \bibnamefont {Lo}},\ }\href@noop {} {\bibfield
  {journal} {\bibinfo  {journal} {New J. Phys.}\ }\textbf {\bibinfo {volume}
  {15}},\ \bibinfo {pages} {113007} (\bibinfo {year} {2013})}\BibitemShut
  {NoStop}%
\bibitem [{\citenamefont {Ma}(2008)}]{ma2008quantum}%
  \BibitemOpen
  \bibfield  {author} {\bibinfo {author} {\bibfnamefont {X.}~\bibnamefont
  {Ma}},\ }\href@noop {} {\bibfield  {journal} {\bibinfo  {journal}
  {arXiv:0808.1385v1}\ } (\bibinfo {year} {2008})}\BibitemShut {NoStop}%
\bibitem [{\citenamefont {Yuan}\ \emph {et~al.}(2016)\citenamefont {Yuan},
  \citenamefont {Zhang}, \citenamefont {L\"utkenhaus},\ and\ \citenamefont
  {Ma}}]{yuan2016simulating}%
  \BibitemOpen
  \bibfield  {author} {\bibinfo {author} {\bibfnamefont {X.}~\bibnamefont
  {Yuan}}, \bibinfo {author} {\bibfnamefont {Z.}~\bibnamefont {Zhang}},
  \bibinfo {author} {\bibfnamefont {N.}~\bibnamefont {L\"utkenhaus}}, \ and\
  \bibinfo {author} {\bibfnamefont {X.}~\bibnamefont {Ma}},\ }\href@noop {}
  {\bibfield  {journal} {\bibinfo  {journal} {Phys. Rev. A}\ }\textbf {\bibinfo
  {volume} {94}},\ \bibinfo {pages} {062305} (\bibinfo {year}
  {2016})}\BibitemShut {NoStop}%
\end{thebibliography}
%%%%%%%%%%%%%%%%%%%%%%%%%%%%%%%%%%%%%%%%

%

%\input{.bbl}
\end{document}